\documentclass[twocolumn,showpacs]{revtex4}
\usepackage{times,xspace}
\usepackage{amsbsy,amssymb,amsmath,bm}
\usepackage{graphicx,color,epsfig,rotate}
\usepackage{fancyhdr}

\def\bbbc{{\mathchoice {\setbox0=\hbox{$\displaystyle\rm C$}\hbox{\hbox
to0pt{\kern0.4\wd0\vrule height0.9\ht0\hss}\box0}}
{\setbox0=\hbox{$\textstyle\rm C$}\hbox{\hbox
to0pt{\kern0.4\wd0\vrule height0.9\ht0\hss}\box0}}
{\setbox0=\hbox{$\scriptstyle\rm C$}\hbox{\hbox
to0pt{\kern0.4\wd0\vrule height0.9\ht0\hss}\box0}}
{\setbox0=\hbox{$\scriptscriptstyle\rm C$}\hbox{\hbox
to0pt{\kern0.4\wd0\vrule height0.9\ht0\hss}\box0}}}}

\newcommand{\ignore}[1]{}
\newcommand{\mComment}[1]{}
\newcommand{\gComment}[1]{}
\newcommand{\jComment}[1]{}
\newcommand{\rComment}[1]{}
\newcommand{\lComment}[1]{}

\renewcommand{\mComment}[1]{\textcolor{blue}{Manny: #1}}
\renewcommand{\gComment}[1]{\textcolor{red}{Gerardo: #1}}
\renewcommand{\jComment}[1]{\textcolor{green}{Jim: #1}}
\renewcommand{\rComment}[1]{\textcolor{magenta}{Ray: #1}}
\renewcommand{\lComment}[1]{\textcolor{purple}{Rolando: #1}}

\begin{document}

\title{Spin and Spin-Wave Dynamics in Josephson Junctions}
\author{Zohar Nussinov$^{1,2}$, Alexander Shnirman$^{3}$, 
Daniel P. Arovas$^{4}$, 
Alexander V. Balatsky$^{1}$, and Jian Xin Zhu$^{1}$}
\address{$^{1}$Theoretical Division,
Los Alamos National Laboratory, Los Alamos, NM 87545, USA}
\address{$^{2}$ Department of Physics, Washington University, St. Louis, 
MO 63160-4899, USA}
\address{$^{3}$  Institut f\"ur Theoretische Festk\"orperphysik,
Universit\"at Karlsruhe,
D-76128 Karlsruhe, Germany}  
\address{$^{4}$ Department of Physics, University of California,
San-Diego, La Jolla, CA 92093, USA}

\date{Received \today }

\begin{abstract}

We extend the Keldysh formulation 
to quantum spin systems and derive exact equations
of motion. This allows us to explore the dynamics of
single spins and of ferromagnets when 
these are inserted between superconducting
leads. Several new effects are reported.
Chief amongst these are nutations
of single S=1/2 spins in Josephson junctions.
These nutations are triggered by the superconducting pairing
correlations in the leads. Similarly, we find that
on rather universal grounds, magnets
display unconventional spin wave dynamics
when placed in Josephson junctions. These lead 
to modifications in the tunneling current.

\end{abstract}

\pacs{71.27.+a, 71.28.+d, 77.80.-e}

\maketitle

\section{Introduction} 

There is a growing interest in a number of techniques that allow
detection and manipulation of a single spin. A
partial list includes optical detection of electron spin resonance
(ESR) in a single molecule~\cite{Koehler93}, tunneling through a
quantum dot~\cite{Engel01}, and, more recently, the ESR-scanning
tunneling microscopy (ESR-STM) technique~\cite{Mana89,Durkan02}.
Interest in ESR-STM lies in the potential of detection and
manipulation of a single spin~\cite{Manoharan02,Balatsky02}- an ability 
which is crucial to spintronics and quantum information processing. 
Much work also addressed 
coupling, feedback effects, and decoherence in a coupled
electronic-vibrational systems, such as nanomechanical oscillators
and local vibrational modes~\cite{Mozy02b}.
In particular, spintronic and quantum computing applications
greatly intensified interest in Josephson junctions.
In a previous publication \cite{nut03}, four of us studied
the effect of the supercurrent on a macroscopic spin cluster 
(of spin $S \gg 1$)
precessing in the presence of a magnetic field when placed
in a Josephson junction to find new spin dynamics. 
In \cite{bulaevskii}, these systems were examined anew 
wherein variations in the DC current 
were predicted for all systems harboring a spin of any finite size
$S$. In the current article, we complement
\cite{nut03} by studying, for the first time, the explicit dynamics of 
{\em single quantum $S=1/2$ spins} in Josephson junctions
to find new intriguing dynamical effects 
for which we provide quantitative expressions.
The single spin ($S=1/2$) dynamics which we 
study here differs significantly from the 
the large magnetic cluster ($S \gg 1$) dynamics 
studied in \cite{nut03}. In the current 
article, we further examine spin wave dynamics in ferromagnets
when placed in Josephson junctions. 

The analysis of spins embedded in Josephson junctions has a
long and rich history.
Early on, Kulik~\cite{Kulik_JETP66} argued that spin flip processes in
tunnel barriers reduce the critical Josephson current
as compared to the Ambegaokar-Baratoff
limit~\cite{Ambegaokar_Baratoff}. More than a decade later,
Bulaevskii et al.~\cite{Bulaevskii_pi_junction} conjectured that
$\pi$-junctions may form if spin flip processes dominate.
The competition between the Kondo effect and the superconductivity was
elucidated in \cite{Glazman_Matveev_SC_Kondo}. A nice review of 
experimental works on certain aspects of 
magnetic nanoparticles in Josephson junctions
is found in \cite{wolfgang}. Transport
properties formed the central core of many
pioneering works, while spin
dynamics was relegated a relatively trivial secondary role.
In the current article, we report on exact new non-stationary spin dynamics
and illustrate how a quantum $S=1/2$ spin is affected
by the Josephson current. As a consequence of
the Josephson current, spins exhibit non-planar
precessions while subject to the external magnetic field.
As well known, a single quantum spin in a magnetic field 
exhibits circular Larmor precession 
about the direction of the field.
As we report here, when the spin
is further embedded between two superconducting leads, quantum 
pairing correlations lead to new out-of-plane
longitudinal motion, much alike that displayed by a classical mechanical
top will arise.
We term this effect the {\em Josephson nutation}. Similar effects occur
when a ferromagnetic slab is placed between two superconducting leads.
We outline how transport is, in turn, modulated by this rather 
unusual spin dynamics. 
The coupling of the spin to the supercurrent leads to an effective non-local 
in time interaction of the single spin with itself. 
Keldysh contour calculations illustrate 
that a non-local
in time single fermion action is also found in situations wherein
the single spin is replaced by an Anderson
impurity~\cite{avishai}. As well known, in the limit of small
hopping amplitudes to and from an Anderson impurity, the impurity
attains a Kondo like character much like that of the single spin
which is the focus of our attention. Here we consider the origin
of this rather generic non-locality in time present in the
dynamics of a Josephson junction. En route to deriving this new spin 
dynamics we illustrate that even in the presence of
non-local in time interactions, certain variants of
the classical equations of motion become trivially exact by virtue of
compactness of the spin variables. An elaborate extension of
these ideas will be detailed elsewhere \cite{mns}.

\section{Outline of the article}
The main goal of the current publication
is to report on the spin and spin wave dynamics
(of single spins and of magnetic systems, respectively) in
Josephson junctions. 

To achieve this aim, we will initially
(in Sections(\ref{exact},\ref{keldrot}))
extend the non-equilibrium Keldysh formalism to address
these problems. In Section(\ref{exact}),
we illustrate that even in the presence
of effective non-local of time interactions of
a spin with itself (such as those borne
by the interaction of a single 
spin with a Josephson current),
the equations of motion undergo
a trivial modification. In Section(\ref{keldrot}),
we rewrite these equations within the standard
Keldysh basis best suited for non-equilibrium 
problems. Sections(\ref{exact},
\ref{keldrot}) may be seen as 
independent extensions of basic facets
of the non-equilibrium Keldysh 
formalism for a spin 
system.

In Section(\ref{single_spin}), we apply 
the rather general formalism developed
in Sections(\ref{exact},\ref{keldrot})
to the specific problem of a single $S=1/2$
spin in a Josephson junction (with 
a time independent potential difference
between the two superconducting leads).
We start, in subsection(\ref{system}),
in writing down the relevant Hamiltonian
of such a Josephson junction harboring 
a single spin. In subsection(\ref{scales}),
we briefly highlight the natural time
scales in the problem- which will indeed
come to the fore in the detailed solution which 
we will later expose. In the all-important
subsection(\ref{eff_action}), we highlight
the origin of the effective non-local
in time interactions of the 
spin with itself. Here, we integrate 
out the lead electrons to 
find the effective spin only 
action harboring such 
non-local in time interactions.
These non-trivial interactions
are the reason that we needed
to develop and extend 
(Sections (\ref{exact},\ref{keldrot}))
the Keldysh formalism 
to a very general spin system
with such interactions.
In subsection(\ref{eom.}),
we invoke the results of
Sections(\ref{exact},\ref{keldrot})
to the resultant effective
spin-only action of Section(\ref{eff_action})
to write down the equations of motion
for the spin. In subsection(\ref{exact_soln.}),
we solve these equations of motion 
to lowest order in the spin-dependent tunneling
amplitude. Detailed technical aspects of the
solution on which subsection(\ref{exact_soln.})
dutifully relies on have been relegated to 
appendices B and C. The perturbative 
solution to the equations of motion-
the final equations of
subsection(\ref{exact_soln.})- form one of
the main core results of the current publication.
In subsection(\ref{consequences}), we 
examine the physical meaning 
of this solution of the single
spin problem to unearth several
new predictions for this $S=1/2$ system. 
In this subsection, we aim
to further arm the reader with an
intuitive understanding for the physical
origin of these new effects.
Some of these predicted effects 
(and our prediction of nutation in particular)
are highlighted in Fig.(\ref{FIG:RIGID}).
In subsection(\ref{retS}),
we examine the behavior of the 
system for a single spin of
magnitude $S > 1/2$. In the 
large $S \gg 1$ limit, we
recover our very different 
semi-classical spin ($S \to \infty$) 
results of \cite{nut03}.

Next, in Section(\ref{AC_effects.}), we discuss
a variation of the single spin problem
wherein an AC voltage bias is applied
across the Josephson junction. Our
main result are the predictions
of specific time dependent
spin dynamics displaying 
an infinite number of
harmonics and new DC
lock-in effects. The
predicted supercurrent 
in this system is also 
discussed.

In Section(\ref{ferroS}), we 
examine the problem of
a ferromagnet in a Josephson
junction. In the spin-wave
approximation, we find that 
each spin-wave mode displays some of the unusual
effects predicted in subsection(\ref{exact_soln.},
\ref{consequences}) for the single spin problem.
The predicted spin wave dynamics and 
associated transport (current), are
furnished. In Section(\ref{Geometry.}), we
discuss simple extensions of our 
results to other systems
generated by a trivial
change of geometry wherein
at least one of the superconductors
forming the Josephson junction
is replaced by a planar superconductor.
In Section(\ref{Large_S.}), we 
write down the $S \gg 1$ equations
of motion for general magnets
and antiferromagnetic chains.
The non-uniform temporal evolution
of each of the spin-waves is 
highlighted in the resultant
solution.  We conclude the main text,
in Section(\ref{conc.}),
by highlighting our conclusions.

In Appendix A, we briefly discuss
several experimental manifestations
of our effect and highlight a
proposed experiment which 
may verify our predictions.

\section{Exact Spin-1/2 Equations of Motion on Keldysh contours}
\label{exact}

\begin{figure}
\centerline{\psfig{figure=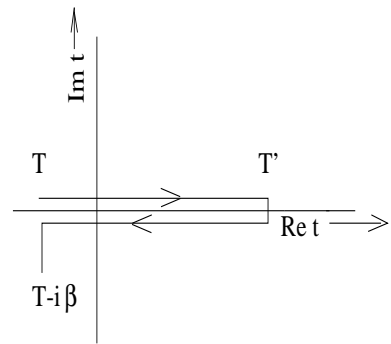,height=4.5cm,width=5cm,angle=0}}
\caption{The standard Keldysh contour.
The times $T$ and $T^{\prime}$ are taken to
be $-\infty$ and $\infty$ respectively. The form of this
contour will be heavily employed in our work when time
ordering various spin products.} \label{FIG:keldysh1}
\end{figure}

We start by deriving the equations of motion
for a very general spin-1/2 system having
two (or more) local and non-local spin-spin
interactions at different times. In this work 
we employ the non-equilibrium Keldysh technique.
Within this framework, the spin operators
on both up and down portions of the (Keldysh) contour
of Fig. 1 are normalized
and satisfy $[\vec{S}_{u}(t), \vec{S}_{d}(t')] =0$. 
In what briefly follows, we will dispense with
operator formulations and employ a
path integral representation. Towards
this end, our working horses will be 
the $CP_{1}$ spin coherent 
variables ($z$) \cite{WZW,perel}
wherein the spins are represented by
\begin{eqnarray}
\vec{S} = S z^{*}_{a} \vec{\sigma}_{ab} z_{b}
\label{CP1.}
\end{eqnarray} 
(with $S$ the spin magnitude). Here and throughout, we set $\hbar=1$. 
In Eq.(\ref{CP1.}), $a,b = \uparrow, \downarrow$ and we assume
an implied summation over repeated indices. The vectors
$\vec{\sigma}_{ab}$ are the $ab$ components
of the three Pauli matrices.
The components $z_{a= \uparrow, \downarrow}$ code 
for a two component 
complex spinor subject to the normalization constraint, 
$|z_{\uparrow}|^{2} + |z_{\downarrow}|^{2}=1$.
By glancing at Eq.(\ref{CP1.}), we note that 
a knowledge of $\vec{S}$ specifies the two component spinor
$z$ only up to a global multiplicative phase.

As well appreciated, in a spin coherent 
basis, the Berry phases associated with the spin coherent states
are the net area of the spherical triangle spanned 
by the spin as it moves on the Bloch sphere.
The latter may be expressed in the $CP_{1}$ 
basis as $S_{Berry} =   i \int dt  \sum_{a} z_{a}^{*} \partial_{t}
z_{a}$ \cite{WZW,perel}. For the benefit of readers
unfamiliar with this formalism, we provide in \cite{explain_CP1} 
a quick derivation for this form of the Berry phase.

We now assume the action contains
the single spin term $- \eta_{a} \int dt \vec{S}_{a} \cdot \vec{h}$
describing a single spin in an external magnetic field set by $\vec{h}$. 
The parity $\eta_{a} = \pm 1$ is fixed by the direction of the
contour- $\eta_{up} = 1, \eta_{down} = -1$.
We further include a non-local in time spin interaction
$ \eta_{a} \eta_{b} \int dt  \int dt^{\prime}
~\overline{K}_{ab}(t,t^{\prime}) \vec{S}_{a}(t) \cdot
\vec{S}_{b}(t^{\prime})$. The kernels $\overline{K}_{ab}$ encapsulate
non-local temporal dependence. The generalization to higher order
terms is straightforward and leads to no qualitative change.
With the Berry phase included, the general action
\begin{eqnarray}
S = 2 iS \eta_{a} \int dt z_{a}^{*} \partial_{t} z_{a}
- S \int dt \eta_{a} \vec{h}_{a} \cdot z_{a}^{*}
\vec{\sigma} z_{a} \nonumber
\\ +S^{2} \eta_{a} \eta_{b} \int dt \int dt^{\prime}
\overline{K}_{ab}(t,t^{\prime}) z_{a}^{*}(t) \vec{\sigma} z_{a}(t)
\cdot z_{b}^{*}(t^{\prime}) \vec{\sigma} z_{b}(t^{\prime}).
\label{Sg}
\end{eqnarray}
Varying the action,
\begin{eqnarray}
\frac{\delta S}{\delta z^{*}_{a \nu}(t)} =
S \eta_{a} \Big(2i \partial_{t} z_{a \nu}(t) - \vec{h} \cdot
\vec{\sigma}_{\nu \gamma} z_{a \gamma}(t) \nonumber
\\ + \eta_{b} \int dt^{\prime} K_{ab}(t,t^{\prime})
z_{b \gamma}^{*}(t^{\prime}) \vec{\sigma}_{\gamma \delta}
z_{b \delta}(t^{\prime})
\cdot \vec{\sigma}_{\nu \gamma} z_{a \gamma}(t) \Big) \nonumber
\\ \equiv S \eta_{a} \Big( 2i \partial_{t} z_{a \nu}(t) - \vec{H}(t)
\cdot \vec{\sigma}_{\nu \gamma} z_{a \gamma}(t) \Big).
\label{fd}
\end{eqnarray}
Here,
\begin{eqnarray}
K_{ab}(t,t^{\prime}) \equiv \overline{K}_{ab}(t,t^{\prime})
+ \overline{K}_{ba}(t^{\prime}, t) \nonumber
\\
\vec{H}(t) \equiv \vec{h} + S \eta_{a} \int dt^{\prime}
K_{ab}(t,t^{\prime}) z_{b \gamma}^{*}(t^{\prime})
\vec{\sigma}_{\gamma \delta} z_{b \delta}(t^{\prime}).
\label{KH}
\end{eqnarray}

Next, we briefly generalize Ehrenfest's theorem
to situations such as the one of relevance
here where a non-local in time action is present. 
A full discussion of this theorem 
for general systems will be presented elsewhere
\cite{mns}. In what follows, the expectation value of any quantity 
${\cal{A}}$ evaluated with the action $S$ is denoted by
\begin{eqnarray}
\langle {\cal{A}} \rangle_{S} \equiv \frac{1}{Z} \int Dz Dz^{*} \delta(|z|^{2}-1) 
{\cal{A}} e^{iS},
\end{eqnarray}
with $Z =  \int Dz Dz^{*} \delta(|z|^{2}-1)  e^{iS}$
the associated partition function. Similar definitions
apply, with a trivial replacement of
the measure when the action is a functional of one of more 
real fields $\{x_{\alpha}(t)\}$. In the current context, $x_{\alpha}$ code
for the real or imaginary parts of the complex spinor 
components $z$. Next, we note that for any cyclic coordinate $x$, 
the expectation value of the variational derivative, 
\begin{eqnarray}
\langle \frac{\delta S}{\delta x} \rangle_{S} =
\frac{-i}{Z} [e^{iS}]_{x_{i}(t)}^{x_{f}(t)}=0.
\label{Ehrenfest}
\end{eqnarray}
In the above, by the compactness of $x$,
in integrating all possible trajectories $x(t)$, the initial
and final trajectories are equal $x_{i}(t) = x_{f}(t)$. 
This in turn lead to the vanishing expectation value
given in Eq.(\ref{Ehrenfest})
for all non-singular actions.
Analogously, this result follows by noting that for compact
coordinates, the transformation 
[$x(t) \to x(t) + \delta x(t)$], with any $\delta x(t)$ leads to 
no change to the value of $Z$- the range of integration in 
$Z= \int Dx~ e^{iS}$ is unchanged. This, in turn, mandates 
that $\langle \frac{\delta  S}{\delta x} \rangle_{S} =0$ \cite{mns}.
Next, we consider ${\cal{A}}^{i} \equiv z^{*} \sigma^{i} 
\frac{\delta S}{\delta z^{*}}$ and explicitly illustrate that its expectation
value vanishes, $\langle {\cal{A}}^{i} \rangle  = 0$. 
To this end, we write the spinors, longhand, in terms of 
real and imaginary components, 
$z^{*} = (z^{1}_{Re} - i z^{1}_{Im}~ z^{2}_{Re} - i z_{Im}^{2})$
and the measure $DzDz^{*} \delta(|z|^{2}-1) = Dz^{1}_{Re} Dz^{1}_{Im} 
Dz^{2}_{Re} Dz^{2}_{Im} \delta(|z^{1}_{Re}|^{2} + |z^{1}_{Im}|^{2} +
|z^{2}_{Re}|^{2} + |z^{2}_{Im}|^{2} -1)$. Here and in what briefly follows
we suppress a uniform Keldysh contour index. The expectation value
$\langle A^{i} \rangle$ for each value of the spin index, $i=x,y,z$, is an
integral over bilinears in $z$ and hence amounts
to a sum of integrals of the type 
\begin{eqnarray}
I_{\alpha \beta} \equiv 
\int Dz^{1}_{Re} Dz^{1}_{Im} 
Dz^{2}_{Re} Dz^{2}_{Im} \nonumber
\\ \delta(|z^{1}_{Re}|^{2} + |z^{1}_{Im}|^{2} +
|z^{2}_{Re}|^{2} + |z^{2}_{Im}|^{2} -1) 
z_{\alpha} \frac{\delta S}{\delta z_{\beta}} e^{iS}.
\end{eqnarray}
Here, the indices $\alpha$ and $\beta$ span the four possible values
$(1 ~Re, 1 ~Im, 2 ~Re, 2~Im)$. An immediate consequence of the vanishing 
of the expectation value  $\langle \frac{\delta S}{\delta x} \rangle$
for any cyclic coordinate $x$ is that all integrals 
of the form $I_{\alpha \neq \beta}$ vanish. An inspection 
of $\langle {\cal{A}}^{i} \rangle$ reveals
that the contributions of all integrals of the type 
$I_{\alpha \alpha}$ cancel identically
when $i=z$ (the only place where integrals
of the type $I_{\alpha = \beta}$
appear). Similarly, $\langle {\cal{A}}^{i \dagger} \rangle
= \langle \frac{\delta S}{\delta z(t)} 
\sigma^{i} z \rangle = 0$. The
vanishing $\langle {\cal{A}}^{i} \rangle = 
\langle {\cal{A}}^{i \dagger} \rangle =0$
imply that their difference, 
\begin{eqnarray}
0= \langle [\frac{\delta S}{\delta z_{a \sigma}(t)} 
\sigma^{j}_{\sigma \sigma^{\prime}} z^{a \sigma^{\prime}}
- z^{*}_{a \sigma} \sigma^{j}_{\sigma \sigma^{\prime}}
\frac{\delta S}{\delta z^{*}_{a \sigma^{\prime}}(t)}] \rangle_{S}, 
\label{identity}
\end{eqnarray}
where the Keldysh contour index ($a$) is reinstated.
 
Next, we explicitly insert Eq.(\ref{fd}) into Eq.(\ref{identity}).  
As a consequence of the SU(2) algebra
of the Pauli matrices, we find that for each Keldysh contour index 
$a=$ top/bottom,
\begin{eqnarray}
\langle \frac{\partial \vec{S}_{a}}{\partial t} \rangle_{S} = - \langle
\vec{H} \times \vec{S}_{a} \rangle_{S}.
\label{ud}
\end{eqnarray}
Eq.(\ref{ud})
is none other than the equation of motion for precession
of the spin $\vec{S}$ in the instantaneous
field given by $\vec{H}$ of Eq.(\ref{KH}). We find that such 
classical equations of motion for a nonlocal in time
action are exact in the quantum arena.
[For affectionados of parafermion methods, we briefly note as an aside
that although throughout we employed the bosonic 
spin coherent path integral representation,
a similar result follows if the spinors $z$ were Grassmann variables
(a net even number of permutations of the spinor coordinates
are involved in proving Eq.(\ref{ud})).] 
The bulk of the paper will be devoted to a solution of
Eq.(\ref{ud}) for different realizations
of a Josephson junction system.

We will momentarily dispense with the Keldysh
contour indices. Due to the commutation relations 
$\vec{S} \times \vec{S} = i \vec{S}$, 
although the field $\vec{H}$ contains a 
piece which is linear in $\vec{S}$, the planar components of 
Eq.(\ref{ud}), may be reduced for certain 
problems to a linear
equation in planar spin components 
$\langle S_{i} \rangle$ ($i=x,y$)
which then must have the solution 
\begin{eqnarray}
\langle S_{i}(t) \rangle = U_{ij}(t) 
\langle S_{j}(0) \rangle.
\label{SF}
\end{eqnarray} 
We now invoke symmetry constraints.
An external magnetic field $\vec{h}$ in the action 
(Eq.(\ref{Sg})) lifts the $SU(2)$ spin rotational symmetry of the 
free spin leading in turn to a lower $U(1)$ symmetry 
of rotations about the external magnetic field
axis. Such a symmetry is trivially encapsulated by
the operator $R^{z}(\theta)$ rotating $\langle \vec{S} \rangle$
by an angle $\theta$ about the z (or magnetic field) axis. 
As a consequence,  
the evolution operator $U(t)$ of Eq.(\ref{SF}) must commute with 
$R^{z}(\theta)$. 
This, in turn, dictates that if the solution is in the
form of Eq.(\ref{SF}), then the time evolution 
operator $U(t)$ must have the form 
\begin{eqnarray}
U(t) = \left(
 \begin{array}{cc}
p(t) & q(t)\\
-q(t) & p(t) \\
\label{Uxy}
\end{array}
\right).
\end{eqnarray}
Similarly, due the azimuthal rotational symmetry encapsulated
by $R^{z}(\theta)$, the expectation value
$\langle S_{z}(t) \rangle$ 
must be independent of $\langle S_{x}(0) \rangle$ and $ \langle 
S_{y}(0) \rangle$. 
This form will indeed be borne out for our
full Keldysh problem.

\section{The Keldysh basis equations of motion}
\label{keldrot}

Within the non-equilibrium Keldysh formalism it is often advantageous
to apply a simple linear transformation from the basis
of up and down contour fields to the symmetric and
antisymmetric linear combination of these
fields. E.g., for the spin
\begin{eqnarray}
\vec{S}_{cl} \equiv
\frac{1}{2}(\vec{S}_{up} + \vec{S}_{down}), \nonumber
\\ \vec{S}_{qu} \equiv (\vec{S}_{up} - \vec{S}_{down}).
\label{rotation}
\end{eqnarray}
The utility of this basis has its roots in the natural form
for the various correlation functions- all simply related to
the advanced, retarded, and ``Keldysh'' correlators.
The subscripts ``cl'' and ``qu'' of Eq.(\ref{rotation}) coding for 
``classical'' and ``quantum'' suggest
an intimate relation to classical and quantum
Langevin like dynamics. We refer the uninitiated reader
to excellent texts such as \cite{das}, \cite{alex}
where the origin of this link is explored in depth.
In Eq.(\ref{rotation}) we trivially generalize this
change of basis to quantum spin systems.
In this basis, when taken as operators
in Eq.(\ref{rotation}) [prior to a passage
to a path integral representation], 
the spins no longer obey canonical 
commutation relations  the spins no longer obey canonical commutations
relations (e.g., $[\vec{S}_{qu},\vec{S}_{cl}] \neq 0$)
and are no longer normalized $(\vec{S}_{up} \pm \vec{S}_{down}$
may correspond to
a spin-triplet, $S=1$, or to a spin singlet, $S=0$).
Thus, we may not directly employ the $CP_{1}$ representation in this basis.
For the current purposes, the equations of motion
in this basis may be derived from Eq.(\ref{ud})
for the up and down contour spins,
\begin{eqnarray}
0 = \langle \frac{d}{dt} S_{cl}^{k} + (\vec{h} \times \vec{S}_{cl})_{k}
+ \int dt_{2} \epsilon_{ijk} 
(S^{j}_{cl}(t) S^{j}_{qu}(t))
\nonumber
\\
\left(
 \begin{array}{cc}
 \frac{K_{uu}+K_{ud}- K_{du} - K_{dd}}{2} & 
\frac{K_{uu} - K_{ud} - K_{du} + K_{dd}}{4} \\   
\frac{K_{uu}+ K_{ud} + K_{du} + K_{dd}}{4} & 
\frac{K_{uu} - K_{ud} + K_{du} - K_{dd}}{8} 
\end{array}
\right) \nonumber
\\
\left(
\begin{array}{c}
S_{cl}^{i}(t_{2}) \\
S_{qu}^{i}(t_{2})
\end{array}
\right) \rangle_{S},
\label{exactcl}
\end{eqnarray}
and 
\begin{eqnarray}
0 = \langle \frac{d}{dt} S_{qu}^{k} + (\vec{h} \times \vec{S}_{qu})_{k}
+ \int dt_{2} \epsilon_{ijk} 
(S^{j}_{cl}(t) S^{j}_{qu}(t)) \nonumber
\\
\left(
 \begin{array}{cc}
 K_{uu} + K_{ud} + K_{du} + K_{dd} & 
\frac{K_{uu} - K_{ud} + K_{du} - K_{dd}}{2} \\   
\frac{K_{uu} + K_{ud} - K_{du} - K_{dd}}{2} & 
\frac{K_{uu} - K_{ud} - K_{du} + K_{dd}}{4} 
\end{array}
\right) \nonumber
\\
\left(
\begin{array}{c}
S_{cl}^{i}(t_{2}) \\
S_{qu}^{i}(t_{2})
\end{array}
\right) \rangle_{S}.
\end{eqnarray}
An average over $\exp[iS]$ is implicit in  $\langle
~ \rangle_{S}$. As emphasized
earlier, these are not merely saddle point
equations but are rather exact. 
In the above, although the time arguments were 
not explicitly written down, $K_{\alpha \beta}$ 
serves as a shorthand for $K_{\alpha \beta}(t,t_{2})$.

\section{Single Spin Dynamics in a Josephson Junction}
\label{single_spin}

\subsection{The system}
\label{system}

\begin{figure}
\centerline{\includegraphics[width=0.55\columnwidth]{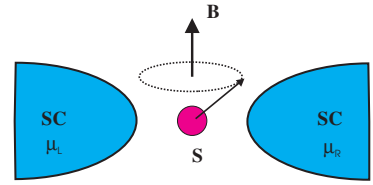}}
\caption{Magnetic spin coupled to two superconducting leads.}
\label{FIG:SETUP}
\end{figure}

Our system is sketched in
Fig.~\ref{FIG:SETUP}. It consists of two identical
ideal superconducting
leads coupled each to a single spin; the
entire system is further subjected to a weak external
magnetic field. In Fig.(\ref{FIG:SETUP}), $\mu_{L,R}$ denote
the chemical potentials of the left and right leads,
${\vec{B}}$ is a weak external magnetic field
along the z-axis,
and ${\vec {S}} = (S_{x}, S_{y},S_{z})$ is
the operator of the localized spin.
The wave-functions of our
system are superpositions of 
the direct product of states of the left contact, the
impurity spin, and the right contact,
\begin{eqnarray}
| \psi \rangle = \sum f_{LSR}  (|\psi_{L} \rangle \otimes 
|\psi_{S} \rangle \otimes |\psi_{R} \rangle).
\label{entangled_eq}
\end{eqnarray}
A tunneling matrix
couples these different states. 
The Hamiltonian of this system
reads
\begin{eqnarray}
{\cal H}={\cal H}_0+{\cal H}_T, \nonumber
\\ 
{\cal H}_0={\cal H}_{L}+{\cal H}_{R}-\mu B_{z}S_z, \nonumber
\\ 
{\cal H}_T=\sum_{\vec{k},\vec{p},\alpha,\alpha'} e^{i\phi/2}\;
c_{R\vec{k}\alpha}^{\dagger}\left[T_0\delta_
{\alpha\alpha'}+T_1\,\mbox{$\vec{\sigma}$}_{\alpha\alpha'}\cdot{\vec
S}\,\right]\,c_{L\vec{p}\alpha'}\nonumber
\\ \!\!\!\!\!\!\! \! \! \! + h.c.
\label{0-T}
\end{eqnarray}
Here, ${\cal H}_{L}$ and ${\cal H}_{R}$ are the Hamiltonians in
the left and right superconducting leads, while $c^{\dagger}_{ik\alpha}$
($c_{ik\alpha}$) creates (annihilates) an electron in the lead a
in the state $\vec{k}$ with spin $\alpha$ in the right/left
lead for $i=L/R$ respectively. 
${\cal H}_{L(R)}=\sum_{k(p);\sigma}\epsilon_{k(p)}
c_{k(p),\sigma}^{\dagger}c_{k(p),\sigma}
 +\frac{1}{2}\sum_{k(p);\sigma,\sigma^{\prime}}
[\Delta_{\sigma\sigma^{\prime}}(k(p)) c_{k(p),\sigma}^{\dagger}
c_{-k(-p),\sigma^{\prime}}^{\dagger} +\mbox{h.c.}]\;,$ where we
denote the electron creation (annihilation) operators in the
left (L) lead by $c_{k\sigma}^{\dagger}$ ($c_{k\sigma}$) while
those in the right (R) lead by $c_{p\sigma}^{\dagger}$
($c_{p\sigma}$). The quantities $k$ ($p$) are momenta, $\sigma$
the spin index, while $\epsilon_{k(p),\sigma}$ and
$\Delta_{\sigma\sigma^{\prime}}(k(p))$ are, respectively, the
single particle energies of conduction electrons, and the pair
potential in the leads. In Eq.(\ref{0-T}), the  
components $\mbox{$\vec{\sigma}$}_{\alpha\alpha'}$
are entries of the three Pauli matrices 
$(\sigma^{x}_{\alpha\alpha'}, \sigma^{y}_{\alpha\alpha'}, 
\sigma^{z}_{\alpha\alpha'})$.
In the current publication we consider 
s-wave symmetry pairing in the superconducting leads.Here, 
$\mu$ is the magnetic moment of the spin. With the spin embedded
in the tunneling barrier, the conduction electron tunneling matrix becomes,
spin-dependent \cite{Balatsky02}
$\hat{T}=[T_0 \hat 1 + T_1 \vec{S} \cdot
\vec{\sigma}_{c}].$
Here $T_0$ is a spin-independent tunneling matrix element and
$T_1$ is a spin-dependent matrix element originating from the
direct exchange coupling $J$ of the conduction electron spin
$\vec{\sigma}_{c}$ to the
localized spin ${\vec{S}}$. Henceforth, 
we will omit the $c$ subscript. We take both tunneling matrix elements
($T_{0}$ and $T_{1}$) to be momentum independent.
This is not a crucial assumption and is merely introduced to
simplify notations. Typically, from the expansion of the work
function for tunneling, $\frac{T_1}{T_0} \sim J/U$, 
where $U$ is the height of 
a spin-independent tunneling barrier~\cite{Zhu_Balatsky}.
A weak external magnetic field $B_{z} \sim 100$ Gauss will not influence
the superconductors and we may
ignore its effect on the leads.
The operator $e^{i\phi/2}$ is the (single electron) number
operator. When the junction is linked to an external environment, the
coupling between the junction and the environment induces fluctuations 
of the superconducting phase difference 
across the junction $(\phi(t))$. 

\subsection{Physical Time Scales} 
\label{scales}

The Josephson junction with
the
 spin has two time scales: (i) The Larmor
 precession frequency of the spin
 $\omega_L = g \mu_B B \equiv h$, where $g, \mu_B$ are the gyromagnetic 
ratio and
 Bohr magneton of the conduction electron, respectively. (ii) The
frequency $\omega_{J} = 2 eV$, with $e$ the electronic charge, 
characterizes the Josephson effect when an external
voltage $V$ is applied across the junction.

\subsection{The Effective Action}
\label{eff_action}

Josephson junctions are necessarily
embedded into external electrical circuits.
This mandates that the dynamics
explicitly depends on the superconducting phase difference
$\phi(t)$ across the junction. 
The evolution operator is given by the real-time path integral
\begin{equation}
Z = \int D\phi D{\vec{S}} \; \exp\left[i S \right] \ .
\label{ZS}
\end{equation}

The net action of Eq.(\ref{ZS}) is given by 
$S= [\mathcal{S_{\rm
circuit}}(\phi)+ \mathcal{S_{\rm spin}}({\vec{S}})+
\mathcal{S_{\rm tunnel}(\phi,{\vec{S}})}]$.
The effective action $\mathcal{S_{\rm tunnel}}$ contribution 
describes the junction itself. If
all external fields are the same on both forward and backward
branches of the Keldysh contour ($K$) then $\mathcal{Z} =
\mbox{Tr}\, T_K\,\exp [-i \oint_K dt H_{T}(t)]  = 1$, where the
trace is over both the electron and the spin degrees of freedom
and $T_{K}$ denotes time ordering along the Keldysh contour.
The label $\oint_K$ denotes integration
along the Keldysh contour as shown in Fig.(\ref{FIG:keldysh1}). 
We first take a partial trace in $\mathcal{Z}$ over the lead
fermions (the bath) to obtain an effective spin action. The
Josephson contribution to the resulting spin action reads
$- \frac{1}{2} \oint_K dt \oint_K dt' \langle
T_K {\cal H}_{T}({\mathbf S(t)},t) {\cal H}_{T}({\mathbf
S(t^{\prime})},t^{\prime}) \rangle$, much in the spirit of
Refs.~\cite{AES_82,Larkin_Ovchinnikov_83,ESA_84}. For brevity, 
we set $A_{\sigma,\sigma^{\prime}} \equiv \sum_{k,p}
c_{k\sigma}^{\dagger} c_{p\sigma^{\prime}}$. The tunneling
Hamiltonian of a phase (voltage) biased junction
\begin{eqnarray}
 {\cal H}_T = [T_0 \delta_{\sigma \sigma'} + T_1 \mathbf{S} \cdot
 \bf{\sigma}_{\sigma \sigma'}]  \nonumber
\\ \times \big(A_{\sigma
\sigma^{\prime}} \exp(i\phi/2) + A_{\sigma \sigma^{\prime}}^\dag
\exp (-i \phi/2)\big)\;. \label{EQ: A1}
 \end{eqnarray}
In the presence of a dc voltage bias, $\phi = 2 eVt$. If $\phi$ is
treated classically (i.e. $\phi$ is the same on the upper and the
lower branches of the Keldysh contour), the contribution $\propto
T_0^2$ to $\delta \mathcal{S}$ vanishes. The mixed contribution
$\propto T_0 T_1$ vanishes due to the singlet spin structure of
the s-wave superconductor. The only surviving contribution reads
\begin{eqnarray}
-\frac{T_1^2}{2} \oint_K dt\oint_K
dt' \left[{\bf S}(t)\cdot \bf{\sigma}_{\alpha \beta}\right] \,
\left[{\bf S}(t') \cdot \bf{\sigma}_{\delta \gamma}\right] \nonumber
\\ \times \left(\langle T_K A_{\alpha \beta}(t) A_{\delta \gamma}(t')
\rangle e^{ i\frac{\phi(t) +\phi(t')}{2}} + (A,\phi\rightarrow
A^{\dag},-\phi)\right) &
 \label{EQ:Seff1}
\end{eqnarray}
where we keep only the Josephson (off-diagonal) terms. The spin
structure simplifies for the s-wave case: \begin{eqnarray}
&& T_1^2 \oint_K dt\oint_K dt' \left[{\bf
S}(t)\cdot {\bf S}(t')\right] [iD(t,t')] \ , \label{EQ:Seff3}
\end{eqnarray}
where the kernel $iD(t,t')$ is dictated 
by $\langle T_K A_{\uparrow\uparrow}(t)
A_{\downarrow\downarrow}(t')\rangle e^{ i\frac{\phi(t)
+\phi(t')}{2}} + (A,\phi\rightarrow A^{\dag},-\phi)$. The operators
$A$ are bilinears in Fermi operators and thus the correlator
$\langle T_K A_{\uparrow\uparrow}(t)
A_{\downarrow\downarrow}(t')\rangle$
will amount to a sum of a product of two terms: 
a product of two normal Green's function $G$ and
a product of two pair correlators $F$. Thus, generalizing 
the known effective tunneling action for a spin-less
junction~\cite{AES_82,Larkin_Ovchinnikov_83,ESA_84} to the
new spin-dependent arena, we obtain
\begin{eqnarray}
\label{EQ:S_EFF} \mathcal{S_{\rm tunnel}}  = - 2 \oint_K dt\oint_K
dt'\, \alpha(t,t')\,\left[T_0^2 + T_1^2 {\bf S}(t)\cdot{\vec
S}(t')\right]\, \nonumber
\\ \cos\left[{\frac{\phi(t) -\phi(t')}{2}}\right]
\nonumber \\ - 2 \oint_K dt\oint_K dt'\,
\beta(t,t')\,\left[T_0^2 - T_1^2 {\vec S}(t)\cdot{\vec
S}(t')\right]\, \nonumber
\\ \cos\left[{\frac{\phi(t) +\phi(t')}{2}}\right] \ ,
\end{eqnarray}
where $i\alpha(t,t')\equiv G(t,t')G(t',t)$ and $i\beta(t,t')\equiv
F(t,t')F^{\dag}(t,t')$. Here, the Green functions
\begin{eqnarray}
&&G(t,t')\equiv -i\sum_\mathbf{k}\,\langle T_{K}
c_{\mathbf{k}\sigma}^{\phantom{\dagger}}(t)
c_{\mathbf{k}\sigma}^{\dagger}(t') \rangle,\\
&&F(t,t')\equiv
-i\sum_\mathbf{k}\,\langle T_{K}
c_{\mathbf{k}\uparrow}(t)c_{\mathbf{-k}\downarrow}(t') \rangle, \\
&&F^{\dag}(t,t') \equiv -i \sum_\mathbf{k} \langle T_{K}\,
c^{\dag}_{\mathbf{k}\uparrow}(t)\, c^{\dag}_{-\mathbf{k}
\downarrow}(t') \rangle
\ .
\end{eqnarray}

We now express the spin action on Keldysh
contour in the basis of coherent states
\begin{equation}
\mathcal{S}_{\rm spin} = - \oint_{K} dt ~ {\vec{h}} \cdot {\vec{S}} +
\mathcal{S}_{WZNW} \ .
\label{wzwn}
\end{equation}
The second,
Wess-Zumino-Novikov-Witten (WZNW), term in Eq.(\ref{wzwn})
depicts the Berry phase
accumulated by the spin which we discussed earlier in the coherent 
spin representation wherein it amounts to a kinetic bilinear- 
the first term of Eq.(\ref{Sg})).
In the calculations that follow we replace the spin measure $DS$ by the 
coherent spin state measure $Dz Dz^{*}$ and rely on our derived exact 
equations
of motion. We now perform the Keldysh rotation of Eq.(\ref{rotation}), 
defining
the values of the spin and the phase variables.
For the superconducting phase, we
introduce (with notations following Refs.~\cite{AES_82,ESA_84})
\begin{equation}
\phi \equiv \frac{1}{2} (\phi_{up} + \phi_{down}) \ \ , 
\ \ \chi \equiv \phi_{up} -
\phi_{down} \ .
\end{equation}
Within the Keldysh framework, the Josephson current is given by 
\begin{eqnarray}
\langle I(t)  \rangle = \frac{2 \pi}{\Phi_{0}} \langle 
\frac{\delta S}{\delta \chi(t)} \rangle,
\label{i(t)}
\end{eqnarray}
with $\Phi_{0}$ the unit fluxon (with full units
restored, $\Phi_{0} = hc/e$ with $c$ the speed of light).
With these definitions in hand, the tunneling part
of the action reads
\begin{equation}
\label{EQ:S_tunnel} \mathcal{S}_{\rm tunnel} =
\mathcal{S}_{\alpha} + \mathcal{S}_{\beta} \ ,
\end{equation}
where the normal (quasi-particle) tunneling part
$\mathcal{S}_{\alpha}$ is expressed via the Green functions
$\alpha^{R}\equiv \theta(t-t')(\alpha^{>} - \alpha^{<})$ and
$\alpha^{K}(\omega) \equiv \alpha^{>} + \alpha^{<}$, where
$i\alpha^{>}(t,t') \equiv G^{>}(t,t')G^{<}(t',t)$ and
$i\alpha^{<}(t,t') \equiv G^{<}(t,t')G^{>}(t',t)$. Similarly the
Josephson-tunneling part $\mathcal{S}_{\beta}$ is expressed via
the off-diagonal Green's functions $\beta^{R}\equiv
\theta(t-t')(\beta^{>} - \beta^{<})$ and $\beta^{K}(\omega) \equiv
\beta^{>} + \beta^{<}$, where $i\beta^{>}(t,t') \equiv
F^{>}(t,t')F^{\dag >}(t,t')$ and $i\beta^{<}(t,t') \equiv
F^{<}(t,t')F^{\dag <}(t,t')$. The pair correlators
$F^{<}(t,t')$ are derived from $F^{>}(t,t^{\prime})$
by the interchange of $t$ with $t^{\prime}$. 
In the current article, we focus on 
the interaction between the supercurrent and the
spin. 

In Eq.~(\ref{EQ:S_EFF}), 
the normal-tunneling part $\mathcal{S}_{\alpha}$ is obtained from
$\mathcal{S}_{\beta}$ by the following substitution:
$\beta^{R/K}(t,t') \rightarrow \alpha^{R/K}(t,t')$, $\phi(t')
\rightarrow -\phi(t')$, and $\chi(t') \rightarrow -\chi(t')$. The
Keldysh terms (those including $\beta^{K}$ and $\alpha^{K}$),
which normally give rise to random Langevin terms (see, e.g.,
Ref.~\cite{ESA_84}) are, in our case, suppressed at temperatures
much lower than the superconducting gap ($T \ll \Delta$), due to
the exponential suppression of the correlators
$\beta^{K}(\omega)$ and $\alpha^{K}(\omega)$ at $\omega <
\Delta$.

To obtain $\beta^{R}$ we start from the Gor'kov Green functions
\begin{eqnarray}
F^{>}(t,t') = -i \sum_k \frac{\Delta}{2 E_k} e^{-iE_k (t-t')},  \nonumber
\\ F^{>\dag}(t,t')  = i \sum_k \frac{\Delta}{2 E_k} e^{-iE_k
(t-t')} \ ,
\end{eqnarray}
where the quasi-particle energy
$E_k\equiv \sqrt{\Delta^2 + \epsilon_k^2}$, with $\epsilon_k$
the free-conduction-electron dispersion in the leads. Putting
all of the pieces together, we find that
\begin{equation}
\beta^{R}(t-t')= - \theta(t-t') \sum_{k,p} \frac{\Delta^2}{2 E_k E_p}
\sin\left[(E_k+E_p)(t-t')\right]\ .
\label{BetaR}
\end{equation}
The kernel $\beta^{R}(t-t')$ decays on (short) time scales
of order ${\cal{O}}(\hbar/\Delta)$.
Similarly,
\begin{eqnarray}
\beta^{K}(t-t^{\prime}) = - i \sum_{k,p} 
\frac{\Delta^2}{2 E_k E_p}
\cos \left[(E_k+E_p)(t-t')\right]\ .
\label{BetaK}
\end{eqnarray}
Henceforth, we will often employ the shorthand $\beta^{R/K}(t,t^{\prime}) 
\equiv \beta^{R/K}(t-t^{\prime})$. 
Looking at Eq.(\ref{BetaK}), we see that the 
Fourier transform $\beta^{K}(\omega)$ vanishes for 
frequencies $\omega < \Delta$. This
is not so for the retarded correlator 
$\beta^{R}$ due to the presence of the theta function.
For now, we ignore the fluctuations in the superconducting
phase and set $\phi_{up}(t) = \phi_{down}(t) = \phi(t) = \omega_{J}t$ 
with $\omega_{J} = 2eV$ (and thus $\chi=0$).
In this, ``classical'', limit
\begin{eqnarray}
S_{tunnel} \simeq  4 \int dt \int dt^{\prime} \beta^{R}(t,t^{\prime})
T_{1}^{2} \vec{S}_{qu}(t) \cdot \vec{S}_{cl}(t^{\prime})  j(t,t^{\prime})
\nonumber
\\
+ \int dt \int dt^{\prime} T_{1}^{2} \beta^{K}(t,t^{\prime})
\vec{S}_{qu}(t)
\cdot \vec{S}_{qu}(t^{\prime}) j(t,t^{\prime}),
\label{Stunnel*}
\end{eqnarray}
with
$j(t,t^{\prime}) \equiv \cos \frac{\phi(t)
+ \phi(t^{\prime})}{2}$.

\subsection{The Equations of Motion}
\label{eom.}

With the action at our disposal, we now write down the 
exact equations of motions and give a
solution, exact to order ${\cal{O}}(T_{1}^{2})$.
Extracting, in the up-down contour basis, 
the coefficients, $\overline{K}_{ab}(t,t^{\prime})$ 
of the $\vec{S}_{a}(t) \cdot \vec{S}_{b}(t^{\prime})$ terms
in Eq.(\ref{Stunnel*}), constructing $K_{ab}(t,t^{\prime})$ from 
Eq.(\ref{KH}), and invoking Eq.(\ref{exactcl}), 
we find
\begin{eqnarray}
0 = \langle \frac{d}{dt} \vec{S}_{cl} + \vec{h} \times \vec{S}_{cl} \nonumber
\\ + 4 |T_{1}|^{2}
\int dt^{\prime} j(t,t^{\prime})
\beta^{R}(t,t^{\prime}) \vec{S}_{cl}(t^{\prime})
\times \vec{S}_{cl}(t) \nonumber
\\ + 2 |T_{1}|^{2} \int dt^{\prime} j(t,t^{\prime})
\beta^{K}(t,t^{\prime})  \vec{S}_{qu}(t^{\prime})
\times \vec{S}_{cl}(t) \nonumber
\\ + |T_{1}|^{2} \int dt^{\prime} j(t,t^{\prime})
\beta^{R}(t,t^{\prime}) \vec{S}_{qu}(t^{\prime}) \times \vec{S}_{qu}(t)
\rangle_{S} \nonumber
\\ \equiv \langle \frac{d}{dt} \vec{S}_{cl} + \vec{h} \times \vec{S}_{cl}
+ \vec{I}_{cl-cl}+ \vec{I}_{qu-cl} + \vec{I}_{qu-qu} 
\rangle_{S}.
\label{class_final}
\end{eqnarray}
The final subscript $S$ serves to remind us that this
is the path integral average computed with the action $S$.
The various subscripts of the integrals $I$ denote 
the terms that they originate from 
(e.g. $I_{cl-cl} = 4 |T_{1}|^{2}
\int dt^{\prime} j(t,t^{\prime})
\beta^{R}(t^{\prime},t) \vec{S}_{cl}(t^{\prime})
\times \vec{S}_{cl}(t)$). 
In Appendices B and C we outline, in detail,
the evaluation of the various terms in Eq.(\ref{class_final}).
We will now solve Eq.(\ref{class_final}) 
to order ${\cal{O}}(T_{1}^{2})$.

\subsection{Spin Dynamics in a Josephson Junction: An Exact Solution
to ${\cal{O}}(T_{1}^{2})$}
\label{exact_soln.}

With all of the ingredients in place, we may now 
solve Eq.(\ref{class_final}) to
determine the spin dynamics to ${\cal{O}}(T_{1}^{2})$. 
Henceforth, we will examine throughout the observable ``classical''
component of the spin $\vec{S}_{cl}$. To make the expressions more
appealing we will dispense with the classical ``cl'' subscript.
Similarly, the action $S$ subscript in all expectation values 
will be omitted as no time ordering subtleties appear below.
We expand the spin as 
\begin{eqnarray}
\langle \vec{S}(t) \rangle = \langle \vec{S}_{0}(t) \rangle + \langle
\delta \vec{S}(t) \rangle.
\label{deltacorrection}
\end{eqnarray}
Here, $\vec{S}_{0}$ is the solution to the (Larmor) problem 
of a single free spin in an external magnetic
field. We computed the integrals borne
by these zeroth order Larmor components in subsection(\ref{integrals}). 
Similarly, $\delta S(t)$ are the 
contributions borne by the retarded and Keldysh
correlations. These corrections will lead to 
higher order contributions in $\langle \vec{I} \rangle$ which 
are irrelevant to our ${\cal{O}}(T_{1}^{2})$ 
solution. We insert Eq.(\ref{deltacorrection}) 
into the equations of motion
(Eqs.(\ref{class_final}))
and retain all terms to order  
${\cal{O}}(|T_{1}|^{2})$.
This trivially leads to 
\begin{eqnarray}
\frac{d}{dt} \langle \delta S_{x} \rangle - \omega_{L} \langle \delta S_{y} 
\rangle  + \langle I_{x} \rangle =0, \nonumber
\\ \frac{d}{dt} \langle \delta S_{y} \rangle + \omega_{L} \langle 
\delta S_{x} \rangle + \langle I_{y} \rangle = 0, \nonumber
\\ \frac{d}{dt} \langle \delta S_{z} \rangle  + \langle I_{z} \rangle = 0.
\label{setxyz}
\end{eqnarray}

Here, $I_{\alpha=x,y,z}$ is the $\alpha$ direction component
of $\langle \vec{I}_{cl-cl}+ \vec{I}_{qu-cl} \rangle$ which was computed
in the previous subsection to order ${\cal{O}}(|T_{1}|^{2})$. 
We see that the integrals $\vec{I}$ play the role
of a driving force. Integrating, we find that
\begin{eqnarray}
\langle \delta S_{z}(t) \rangle = |T_{1}|^{2} (1- \cos \omega_{J} t)
 [\sum_{k,p} \frac{\Delta^{2} 
\omega_{L}}{E_{k} E_{p}(E_{k}+ E_{p})^{3}} \nonumber
\\ + \langle S_{z}(0) \rangle \sum_{k,p} \frac{\Delta^{2}}{E_{k}E_{p} 
(E_{k}+E_{p})^{2}}].
\label{deltaSz}
\end{eqnarray}

Differentiating the equation of motion for $\langle \delta S_{x,y} 
\rangle$ in Eq.(\ref{setxyz}) and inserting the equation of motion
for $\langle \delta S_{y,x} \rangle$ we immediately
obtain the equation of motion of a driven harmonic oscillator.
A simple solution yields
\begin{eqnarray}
\langle \delta S_{x}(t) \rangle 
= c_{1} \cos \omega_{L} t + c_{2} \sin \omega_{L} t \nonumber 
\\ + \sum_{\omega_{n} } (\frac{A_{n}}{\omega_{L}^{2} - \omega_{n}^{2}} 
\cos \omega_{n} t 
+ \frac{B_{n}}{\omega_{L}^{2} - \omega_{n}^{2}} \sin \omega_{n} t)
\label{sxsoln}
\end{eqnarray}
with
\begin{eqnarray}
A_{\omega_{L} + \omega_{LJ}} =  |T_{1}|^{2} \sum_{k,p} 
\frac{\Delta^{2} \langle S_{x}(0) \rangle
  (2 \omega_{L}^{2} + \omega_{J}^{2}
+3 \omega_{L} \omega_{J} )}{2E_{k} E_{p} (E_{k}+E_{p})^{2}} 
 \nonumber
\\ A_{\omega_{L} - \omega_{J} } =  |T_{1}|^{2} \sum_{k,p} 
\frac{\Delta^{2} \langle S_{x}(0) \rangle 
(2 \omega_{L}^{2} + \omega_{J}^{2}-3 \omega_{L} 
\omega_{J}) }{2E_{k} E_{p} (E_{k}+E_{p})^{2}} 
 \nonumber
\\ B_{\omega_{L}+ \omega_{J}} = 
  |T_{1}|^{2} \sum_{k,p} 
\frac{\Delta^{2}
\langle S_{y}(0) \rangle (
2 \omega_{L}^{2} + \omega_{J}^{2}+ 3 \omega_{L} \omega_{J})}
{2E_{k} E_{p} (E_{k}+E_{p})^{2}} , \nonumber
\\ B_{\omega_{L}- \omega_{J}} = 
 |T_{1}|^{2} \sum_{k,p} 
\frac{\Delta^{2}
\langle S_{y}(0) \rangle (2 \omega_{L}^{2} + \omega_{J}^{2}
- 3 \omega_{L} \omega_{J})}{2E_{k} E_{p} (E_{k}+E_{p})^{2}} .
\label{AB}
\end{eqnarray}

All in all, to ${\cal{O}}(T_{1}^{2})$, the evolution of 
the planar spin components can be expressed in the format 
of Eqs.(\ref{SF},\ref{Uxy}) with 
\begin{eqnarray}
p(t) = \cos \omega_{L} t + |T_{1}|^{2} \sum_{k,p} \frac{\Delta^{2}}
{2 E_{k} E_{p} (E_{k}+ E_{p})^{2}} \nonumber
\\ \times \
\Big( \frac{(2 \omega_{L}^{2} + \omega_{J}^{2} + 3 \omega_{L} \omega_{J})
\cos(\omega_{L} + \omega_{J}) t}
{\omega_{L}^{2} - (\omega_{L} + \omega_{J})^{2}} \nonumber
\\ + \frac{(2 \omega_{L}^{2} + \omega_{J}^{2} - 3 \omega_{L} \omega_{J})
\cos(\omega_{L} - \omega_{J}) t}{\omega_{L}^{2} - (\omega_{L} - \omega_{J})^{2}}
\Big),
\label{pform}
\end{eqnarray}
and 
\begin{eqnarray}
q(t) = \sin \omega_{L} t + |T_{1}|^{2} \sum_{k,p} \frac{\Delta^{2}}
{2 E_{k} E_{p} (E_{k}+ E_{p})^{2}} \nonumber
\\ \times \Big( \frac{(2 \omega_{L}^{2} + \omega_{J}^{2} + 3 \omega_{L} 
\omega_{J})
\sin(\omega_{L} + \omega_{J}) t}
{\omega_{L}^{2} - (\omega_{L} + \omega_{J})^{2}} \nonumber
\\
+ \frac{(2 \omega_{L}^{2} + \omega_{J}^{2} - 3 \omega_{L} \omega_{J})
\sin(\omega_{L} - \omega_{J}) t}{\omega_{L}^{2} - (\omega_{L} - \omega_{J})^{2}}
\Big).
\label{qform}
\end{eqnarray}

This concludes our solution 
for the dynamics of a spin in a Josephson
junction. Our analysis throughout
centered on Josephson junctions composed of 
s-wave superconductors (see our starting point
Eq.(\ref{EQ:Seff3})). Slightly different quantitative 
results appear for other pairing symmetries
(allowing, in theory, a determination of the pairing
symmetry from observations of the spin/spin-wave dynamics
and associated effects). The deviations from 
simple Larmor precessions are far stronger for 
triplet (i.e. odd angular momenta) superconductors.

\subsection{Physical Consequences: 
Josephson Nutations and Other Dynamical Effects}
\label{consequences}

We now discuss the physics behind our exact
(to ${\cal{O}}(T_{1}^{2})$) solution. 
Our solution provides testimony (and to new {\em quantitative}
predictions) for several, inter-related, intriguing dynamical effects.
We outline these below.

$\bullet$ {\underline{{\em Josephson Nutations:}}

In any system harboring a continuous rotational 
$U(1)$ symmetry about a certain axis (z),
the orbital angular momentum $L_{z}$
is a constant of motion. Needless to say, the same
trivially holds true for any spin
system in which $[H, S_{z}]=0$
with $H$ the system Hamiltonian.
In the presence of an external magnetic
field along (or defining) the z-axis, as in the Larmor problem,
the Hamiltonian $H = - h S_{z}$ commutes with $S_{z}$
and the longitudinal magnetization $\langle S_{z}(t) \rangle$ is a constant
of motion. In our system with non
local in time interactions triggered by superconducting pair
correlations, such a conservation law no longer holds.
Perusing Eq.(\ref{deltaSz}), we find that 
the spin {\em nutates} above its average value.
This occurrence for the $S=1/2$ is similar
to that reported in \cite{nut03} for macroscopic
spin clusters $S \gg 1$. Here, however, 
the quantum fluctuations are profound for the $S=1/2$ case and 
lead to strong deformations of the nutation shape.
The physical engine behind the nutations
is the small time separation between 
the two tunneling electrons forming
the Cooper pair. As the ``first'' 
electron tunnels through, it exerts
a torque on the spin. A certain time
later (of order $\hbar/ \Delta$ with dimensions restored)  
after the spin $\vec{S}$ has already
revolved a small amount, the opposite
spin member of the Cooper pair tunnels
through and exerts a torque of an opposite
sign on the spin $\vec{S}$. Due to the small
time lag between the two tunneling
electrons, the two opposite sign
torques exerted on $\vec{S}$ by the two
opposite sign spins of the tunneling
singlet do not cancel and lead to a net
effect. This origin is made
evident in the retarded correlations
$\beta^{R}$ which further spark a non-vanishing
driving force $\langle I_{cl-cl} \rangle$ along
the z-axis. Mathematically, all of this
results as the tunneling portion of
the action contains terms which trivially
do not conserve $S_{z}$. In the aftermath, 
this led to an effective
time dependent force acting on $S_{z}$.
Its form may be seen by examining
the integral $\langle I_{z} \rangle$ 
appearing in Eq.(\ref{setxyz}). 
The latter is the z-component of the
integrals $\langle \vec{I}_{cl-cl} \rangle$ 
and $\langle \vec{I}_{qu-cl} \rangle$ 
appearing in Eqs.(\ref{Icl-cl*}, \ref{Iqu-cl*}).
(Needless to say, if both members of the Cooper
pair share the same polarization (as in triplet
superconductors) then a far greater effect results.)

A manifestation of the resulting dynamical effect as 
a consequence of these effective external forces in conventional
(s-wave) Josephson Junctions is vividly seen in Eq.(\ref{deltaSz}). 
We have derived similar expressions via an 
independent density matrix approach \cite{NS}.
An exaggerated schematic of this effect
is provided in Fig.(\ref{FIG:RIGID})
which, qualitatively, is none other than the standard
illustration for classical rigid 
body nutations. We find that such motions 
now appear in the quantum arena for a 
single $S=1/2$ particle! The precise shape of our 
trajectories, however, is markedly different from that
exhibited by classical rigid rotors.

\begin{figure}
\centerline{\includegraphics[width=0.5\columnwidth]{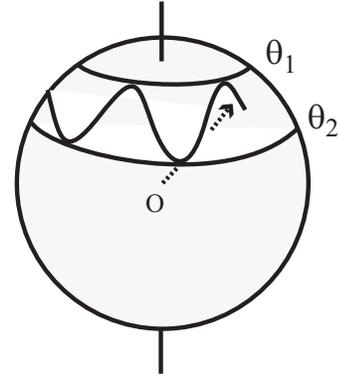}}
\caption{The resulting spin motion on the unit sphere in the
general case. As in the motion of classical spinning top, the spin
exhibits undulations along the polar direction. As a consequence of
entanglement with the tunneling electrons, the magnitude
of the spin is not constant- the spin further ``breaths''
in and out as it nutates.}
\label{FIG:RIGID}
\end{figure}

$\bullet$ {\underline{\em{Spin Contractions and 
Effective Longitudinal Fields:}} 

Glancing at Eq.(\ref{Iqu-cl*}), the reader
will see that the effective $\langle \vec{I}_{qu-cl} \rangle$
can be seen to dilate the spin (the uniform contribution
proportional to $\langle \vec{S}(t) \rangle$
in the second equality of Eq.(\ref{Iqu-cl*}))
and in unison to effectively 
emulate a time dependent 
magnetic field $\vec{\delta h}_{eff} \propto \hat{e}_{z} \cos \phi(t)$ 
along the z-axis in the spin equation of motion,
$d \langle \vec{S} \rangle /dt = ... + \langle \vec{S} \rangle
\times \vec{\delta h}_{eff}$. Both of these effects were
noted in \cite{bulaevskii}. In Eq.(\ref{Iqu-cl*}),
we explicitly see their origin. The uniform contraction 
is triggered by an entanglement of the tunneling electrons 
with our $S=1/2$ particle. We now very briefly elaborate
on the physics of this statement for the benefit of
general readers. The expectation values $\langle \vec{S} \rangle$
amount to weighted sums over all possible states $| \psi \rangle$
(see Eq.(\ref{entangled_eq})). 
In any pure (i.e. unentangled) state
of a spin-1/2 problem, the sum  
$[\langle S_{x} \rangle^{2} + \langle S_{y} \rangle^{2} 
+ \langle S_{z} \rangle^{2}] = 1/4$- the spin expectation
values lie on the Bloch sphere. Entanglement
in a spin-1/2 problem such as ours is marked
by a contraction, $[\langle S_{x} \rangle^{2} + \langle S_{y} \rangle^{2} 
+ \langle S_{z} \rangle^{2}] < S^{2} = 1/4$.  Any single 
spin expectation value within the Bloch sphere, 
$|\langle \psi | \vec{S} | \psi  \rangle| < S$,  
denotes an expectation value computed in a multi-particle state 
$|\psi \rangle$ 
which is entangled. In the case here $|\psi \rangle$ 
spans the single spin and the tunneling electrons. 
Such a time dependent contraction in the norm 
of $\langle \vec{S} \rangle$
relative to the Bloch radius is evident in our exact solution 
of Eqs.(\ref{SF},\ref{Uxy}, \ref{pform}, \ref{qform}, 
\ref{deltacorrection},  \ref{deltaSz}).

$\bullet$ {\underline{\em Nonlinear planar precession:}}

A notable facet of the dynamics given by
the effects discussed above 
are non-uniform planar precessions. We find that 
within the plane transverse to
the applied field direction, the azimuthal
angle describing the spin orientation, 
$\varphi(t) = \tan^{-1}(\langle S_{y}(t) \rangle/\langle S_{x}(t) \rangle)$
is no longer a linear in time. This effect
bears, once again, strong semblance to nutations in classical 
rigid body dynamics. In the Larmor problem
of a free spin in a magnetic field, 
$\varphi(t) = \omega_{L} t$. In our case, the precession
about the applied field direction is no
longer uniform. Its form is encapsulated
in Eqs.(\ref{sxsoln},\ref{AB})
or, alternatively, by Eqs.(\ref{SF}, \ref{Uxy},\ref{pform}.\ref{qform}). 
Once again, mathematically, the origins of this 
effect are rooted in the effective planar (xy) components
of the effective force $\langle \vec{I} \rangle$ appearing
in Eq.(\ref{setxyz}). The explicit form of this 
effective force is given by 
the sum of the two integrals evaluated
in Eqs.(\ref{Icl-cl*}, \ref{Iqu-cl*}) 
and whose origin explicitly lies, once again, 
in the same non-local in time correlations
borne by the superconducting correlations.

In summary, all of the above qualitative findings 
for the problem a single $S=1/2$ spin inserted in
a Josephson junction are made vivid 
in our ${\cal{O}}(T_{1}^{2})$ exact solution.
From Eqs.(\ref{SF},\ref{Uxy},\ref{pform}, \ref{qform}) for 
the planar spin components and from Eqs.(\ref{deltacorrection}, 
\ref{deltaSz}) for the longitudinal spin we clearly see how 
all of these effects come into play.

\section{Retarded Correlations in General Spin $S$ 
Dynamics in a Josephson Junction}
\label{retS}

The equation of motion, Eq.(\ref{ud}),
is valid for all spins $S$. Much
of our formalism follows with
no change. We now examine the 
integrals $\langle \vec{I} \rangle$
in the general spin $S$ problem. 

We find that for a spin of size $S$, 
the integral $\langle \vec{I}_{qu-cl} \rangle$ undergoes
no change relative to its $S=1/2$ form- Eq.(\ref{Iqu-cl*}) remains the same.
The associated physics fleshed out in the second equality 
of Eq.(\ref{Iqu-cl*}) which was described in the previous section
(spin contractions and the presence of an effective longitudinal field)
undergoes no change for the general spin $S$ case.

Next, we evaluate $\langle \vec{I}_{cl-cl} \rangle$. 
For large spins, $S \gg 1$, the product $\langle \vec{S} (t) \times
\vec{S}(t^{\prime}) \rangle$ is well approximated by the
vector product of averages $\langle \vec{S} (t)\rangle \times
\langle \vec{S}(t^{\prime}) \rangle$. Then, the approach of
Ref.~\cite{nut03} is well justified and we can obtain the {\em
Josephson nutations} of a big spin. For any $S$ and $t'>t$ we
obtain
\begin{eqnarray}
&&\langle \vec{S}_{cl}(t^{\prime}) \times \vec{S}_{cl}(t)
\rangle_{S}
\nonumber \\
&=& \frac{1}{2} \Big \langle \big[\{S_{y}(t),S_{z}(t)\}_{+}\,
[\cos\omega_L(t'-t)-1]
\nonumber \\
&-&\{S_{x}(t),S_{z}(t)\}_+\,\sin\omega_L(t'-t)\big]\, 
\hat{e}_{x} \nonumber \\
&+&\big[\{S_{x}(t),S_{z}(t)\}_{+}\,[1-\cos\omega_L(t'-t)]
\nonumber \\
&-&\{S_{y}(t),S_{z}(t)\}_+\,\sin\omega_L(t'-t)\big]\, 
\hat{e}_{y} \nonumber \\
&+&(2S_{x}^2(t)+2S_{y}^2(t))\,\sin\omega_L(t'-t)\,\hat{e}_{z}\ 
\Big \rangle,
\label{SIcl-cl}
\end{eqnarray}
where $\{...\}_+$ denotes an anticommutator. In the $S=1/2$ case, the 
integral stemming from these vectorial product 
correlations is parallel to the z-axis, 
$\langle \vec{I}_{cl-cl} \rangle || \hat{e}_{z}$.
For spins of size $S>1/2$, however, as we see from Eq.(\ref{SIcl-cl}),
the planar ($x,y$) components also come to the fore
and lead to retarding correlation $(\beta^{R})$ effects
in Eq.(\ref{class_final}). Furthermore, the magnitude of
the driving force  $\langle I_{cl-cl;z} \rangle$ along
the z-axis, much unlike the $S=1/2$ case is
time dependent.

For general spins of size $S > 1/2$, both retarded ($\beta^{R}$)
and Keldysh ($\beta^{K}$) are non-zero along any spin 
direction. All of the effects discussed in subsection(\ref{consequences})
are present. 

It is noteworthy to discuss the scaling of all terms 
with the spin size $S$. As evident from Eq.(\ref{SIcl-cl}),
the integral $\langle \vec{I}_{cl-cl} \rangle$ spawned
by retarded correlations scales as $S^{2}$. Similarly, 
as seen from Eq.(\ref{Iqu-cl*}),  whose form holds
for arbitrary $S$, the effective driving force 
$\langle \vec{I}_{qu-cl} \rangle$ generated by 
Keldysh correlations $(\beta^{K}$) scales 
as $S$, i.e. $\langle \vec{I}_{qu-cl} \rangle ~ ~
\alpha ~ ~S$. Thus, for large spins $S \gg 1$,
the retarded contributions overwhelm stochastic Keldysh 
contributions. In the classical limit, $S \to \infty$,
only the retarded contributions remain \cite{nut03}. 

\section{Spin triggered ac effects}
\label{AC_effects.}

Thus far our discussion centered on a Josephson 
junction for a time independent potential difference $V$
(dc voltage bias) between the two superconducting leads for which $\phi(t) = 
\omega_{J} t$ with $\omega_{J} = 2eV$. 

We now briefly sketch matters for an ac voltage bias 
wherein the potential drop is oscillatory in time
and the corresponding
phase difference is $\phi(t) = A \sin \Omega t$.
To make the physics more transparent, we
omit any dc contributions to the voltage
(and thus linear in time contributions to the phase). 
This serves as a caricature of rf driven Josephson junctions
known to exhibit the famous Shapiro steps \cite{shapiro}.

The setup is given by Fig.~\ref{FIG:SETUP} 
for a spin $S=1/2$ particle yet now with 
an ac voltage applied across the junction.
In the sections that follow, we will resume our central 
focus on the constant voltage drop case, $\phi(t) = 
\omega_{J} t$. Only in this short section
do we analyze an applied ac voltage bias.

The calculations for the ac voltage bias case parallel the analysis
of the previous sections. First, we express all
terms by pure harmonics. This is readily achieved by 
relying on the identity
\begin{eqnarray}
e^{iC \sin x} = \sum_{n} J_{n}(C) e^{inx},
\end{eqnarray}
with $\{J_{n}(C)\}$ Bessel functions.
The factor $j(t,t^{\prime})$ of Eq.(\ref{Stunnel*}) and thereafter
now becomes 
\begin{eqnarray}
j(t,t^{\prime}) = \sum_{n,m} J_{n}(\frac{A}{2}) J_{m}(\frac{A}{2}) 
\cos[\Omega(nt+mt^{\prime})].
\end{eqnarray}
The analog of Eq.(\ref{Icl-cl*}) 
for the ac voltage bias case is 
\begin{eqnarray}
\langle \vec{I}_{cl-cl} \rangle_{S} = 
- |T_{1}|^{2} \hat{e}_{z} \sum_{k,p} \frac{\Delta^{2}}{E_{k} E_{p}}
\sum_{n,m} J_{n}(\frac{A}{2}) J_{m}(\frac{A}{2}) \nonumber
\\ \times \frac{2m \Omega \omega_{L} \sin \Omega (n+m)t}{(E_{k}+ E_{p})^{3}}.
\end{eqnarray}
Further resonant (delta function) 
terms make an appearance for $m \ggg 1$.

Similarly, the analog of Eq.(\ref{Iqu-cl*})
reads 
\begin{eqnarray}
\! \! \! \!\! \! \! \!\! \! \! \!\! \! \! \!\! \! \! \!\! \! \! \!\! \! \! \!\! \! \! \!
\langle \vec{I}_{qu-cl} \rangle_{S} =  - |T_{1}|^{2}
\sum_{n,m,k,p} \frac{\ J_{n}(\frac{A}{2}) J_{m}(\frac{A}{2})
\Delta^{2}}{E_{k} E_{p}(E_{k}+ E_{p})^{2}}  \nonumber
\\ \times [ (2 m \Omega \langle S_{x}(t) \rangle \sin (n+m) \Omega t 
- \langle S_{y}(t) \rangle \omega_{L} 
\cos (n+m) \Omega t) \hat{e}_{x} \nonumber
\\ + ( 2 m \Omega \langle S_{y}(t) \rangle  \sin (n+m) \Omega t
+ \langle S_{x}(t) \rangle \omega_{L} \cos(n+m) 
\Omega t) \hat{e}_{y} \nonumber 
\\ \! \! \! \!\! \! \! \!\! \! \! \!\! \! \! \!
+ 2 m \Omega \langle S_{z}(0) \rangle \sin (n+m) \Omega t 
\hat{e}_{z}]. \nonumber
\end{eqnarray}
To ${\cal{O}}(|T_{1}|^{2})$, the nutations are given by
\begin{eqnarray}
\langle \delta S_{z}(t) \rangle 
= |T_{1}|^{2} \sum_{k,p} \frac{\Delta^{2}}{E_{k}E_{p}}
\sum_{n+m \neq 0} J_{n}(\frac{A}{2}) J_{m}(\frac{A}{2})  \nonumber
\\ \times \frac{2m \Omega}{(E_{k}+ E_{p})^{2}}
(\frac{\omega_{L}}{E_{k}+ E_{p}} + \langle S_{z}(0) \rangle) \nonumber
\\ \times \frac{1 - \cos \Omega(n+m) t}{n+m}.
\label{deltaSzAC}
\end{eqnarray}
Higher order effects further enhance this response.
Eq.(\ref{deltaSzAC}) is the ac voltage bias analog 
of Eq.(\ref{deltaSz}) for the 
dc voltage bias case.
The seminal feature of our results is the existence 
of frequencies in the spin dynamics 
of all integer multiples of the voltage bias driving 
ac frequency $\Omega$.  
As the spin alters (via back-action effects) the tunneling
supercurrent, the supercurrent will exhibit
oscillations at all frequencies $\omega_{r} = r \Omega$
with $r$ an integer. Extending the results of
\cite{bulaevskii} to this problem, 
the supercurrent
\begin{eqnarray}
\langle I(t) \rangle = \sin \phi(t)  
\Big[2 \pi^{2} e \rho^{2} \Delta(|T_{0}|^{2} 
- \frac{3}{4} |T_{1}|^{2})  \nonumber
\\ + 4e |T_{1}|^{2} \rho^{2} h 
\langle S_{z}(t) \rangle \Big],
\label{I_shapiro}
\end{eqnarray} 
where $\rho$ is the spin density of states within the leads, with 
the spin given by Eqs.(\ref{deltacorrection}, \ref{deltaSzAC})
with the Larmor $\langle S_{z}(t) \rangle_{0} = \langle S_{z}(0) \rangle$. 

\section{Ferromagnets
In Josephson Junctions}
\label{ferroS}

We now investigate what transpires when 
ferromagnets (instead of a single spin) are immersed
between two s-wave superconductors with a dc bias voltage applied
across the junction (as illustrated in Fig.(\ref{FIG:MAGNET})).
As in the single spin problem, the full problem
involves both the back-action of the spin on the 
phase of the superconductors (ignored
here) and the spin dynamics sparked by the tunneling
current (which we focus on below). Further, for extended
junctions, phasons naturally appear. 
In what follows, we assume that the phases of the two superconductors 
surrounding a single magnetic slab 
have a spatially uniform phase difference $\phi(t)$.
The tunneling action amounts to
a sum over individual tunneling actions
through each of the individual spins labeled
by their sites $\vec{r}$,
\begin{eqnarray}
S_{tunnel} \simeq \nonumber
\\   4 \sum_{\vec{r}} 
\int dt \int dt^{\prime} \beta^{R}(t,t^{\prime})
T_{1}^{2} \vec{S}_{qu}(\vec{r},t) \cdot 
\vec{S}_{cl}(\vec{r},t^{\prime}) j(t,t^{\prime}) \nonumber
\\ 
+ \int dt \int dt^{\prime} T_{1}^{2} \beta^{K}(t,t^{\prime})
\vec{S}_{qu}(\vec{r},t) 
\cdot \vec{S}_{qu}(\vec{r},t^{\prime}) j(t,t^{\prime}).
\nonumber
\end{eqnarray} 
For ferromagnetic spin chains/planes with 
arbitrary exchange constants $J(\vec{r},\vec{r}^{\prime})$,
and scaled external magnetic field $\vec{h}$,
the exact equation of motion reads 
\begin{eqnarray}
0 = \langle \frac{d\vec{S}_{cl}(\vec{r})}{dt} + 
\vec{h} \times \vec{S}_{cl}(\vec{r}) + \vec{I}_{cl-cl;r} + \vec{I}_{qu-cl;r}
\nonumber
\\ + \sum_{\vec{r}^{\prime}} J(\vec{r},\vec{r}^{\prime}) 
\vec{S}_{cl}(\vec{r}^{\prime},t)  \times 
\vec{S}_{cl}(\vec{r},t) \rangle_{S}.
\label{class_final1}
\end{eqnarray} 

\begin{figure}
\centerline{\includegraphics[width=0.8\columnwidth]{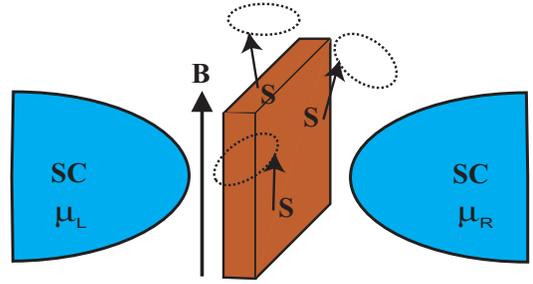}}
\caption{A ferromagnetic slab inserted between two superconducting leads.
The entire system is subjected to a weak external magnetic field $B$.
A schematic of the precessing spins is shown.}
\label{FIG:MAGNET}
\end{figure} 

It is hard not to notice a resemblance between
the single spin problem (Eq.(\ref{class_final}))
and the problem of the ferromagnet (Eq.(\ref{class_final1})).
Indeed, as we will shortly demonstrate the spin wave dynamics
in the ferromagnet within a Josephson junction bears
much in common with the single spin problem 
with the proviso that the various ferromagnetic 
spin waves feel an effective momentum dependent 
magnetic field of strength $h_{eff} = h +S[J(\vec{k})-J(0)]$
with $J(\vec{k})$ the Fourier transform of the 
two spin interaction $J(\vec{r}, \vec{r}^{\prime})$.

The solution proceeds much the same as for the single spin
problem. Henceforth, we discuss
the qualitative physics 
expected. Unlike the precise 
solutions presented till now,
what follows is a quick
qualitative sketch by way
of an analogy. An exact solution
will be detailed elsewhere \cite{long}.

Transforming from spin variables to bosonic operators ($b(\vec{r})$)
at all lattice sites $\vec{r}$,
\cite{DM}
\begin{eqnarray}
S^{+}(\vec{r}) =  b^{\dagger}(\vec{r})\sqrt{2S}, \nonumber
\\ S^{-}(\vec{r}) =  [b(\vec{r}) 
- \frac{1}{2S}  b^{\dagger}(\vec{r}) 
b(\vec{r}) b(\vec{r})]\sqrt{2S}, \nonumber
\\ S_{z} = - S + b^{\dagger}(\vec{r}) b(\vec{r}).
\label{dm}
\end{eqnarray}
Sans the ${\cal{O}}(|T_{1}|^{2}$) 
tunneling part of the action, the action is
quadratic in the bosonic operators and is readily
diagonalized in $\vec{q}$ space. We find that   
the free part of the action 
\begin{eqnarray}
S_{0} = - \int dt \int \frac{d^{d}q}{(2 \pi)^{d}} \{S[J(\vec{q})- J(0)] 
+ h \} \nonumber
\\ \times b^{*}(\vec{q}) b(\vec{q}),
\end{eqnarray} 
with $d$ the dimension of the inserted magnet.
(As the problem is ferromagnetic, $J(0) = \min_{\vec{q}} \{ J(\vec{q})\}$).
Comparing this action to the one appearing in the single 
spin problem, we find that to Gaussian order the spin-wave   
problem is identical to the dynamics of a single spin
with the replacement 
\begin{eqnarray}
h \to h_{eff}(\vec{q}) \equiv \{S[J(\vec{q})- J(0)] 
+ h\}.
\label{heff}
\end{eqnarray} 
The quadratic contribution 
of the ${\cal{O}}(|T_{1}|^{2})$ portion of the action
involving non-local in time correlations
has precisely the same form for both the single 
spin problem and for each mode $\vec{q}$
of the spin-wave problem. 
Thus, the quadratic in $b$, ${\cal{O}}(|T_{1}|^{2})$ corrections
to the spin dynamics are given by Eqs.(\ref{deltaSz},\ref{sxsoln}) 
with the replacement of Eq.(\ref{heff}).

For instance, the above analogy suggests that 
that the net ferromagnetic moment variation 
in $S=1/2$ ferromagnets is
\begin{eqnarray}
\frac{\delta M}{V} \nonumber
\\ =  |T_{1}|^{2} (1- \cos \omega_{J} t)
 [\sum_{k,p} \frac{\Delta^{2} 
\omega_{L}}{E_{k} E_{p}(E_{k}+ E_{p})^{3}} \nonumber
\\ + \frac{M}{V} \sum_{k,p} \frac{\Delta^{2}}{E_{k}E_{p} 
(E_{k}+E_{p})^{2}}],
\label{deltaSzq}
\end{eqnarray}
with $V$ the volume of the magnet
and $M$ its magnetization sans the
supercurrent.
Alternatively, the analysis may 
parallel the derivation of the previous
sections word for word while taking
the unperturbed solution 
(the analogue of the Larmor
solution of Eq.(\ref{Larmoreq}))
to be a spin wave and computing all
corrections to ${\cal{O}}(T_{1}^{2})$.

In the continuum limit,
\begin{eqnarray}
h_{eff}(\vec{q}) = \frac{\rho_{s}}{m_{0}} q^{2} + h,
\end{eqnarray}
with $m_{0} \equiv S/v$ (where $v$ is the volume per site),
the magnetization density of the ground state,
and $\rho_{s}$ the spin stiffness.

This $h \to h_{eff}(\vec{q})$ correspondence 
applies to any property inherited by the 
single spin dynamics in the Josephson
junction. In particular, in \cite{bulaevskii}
it was beautifully shown how spin dynamics may alter
the super-current in the Junction. The  current
may be computed by the likes of Eq.(\ref{i(t)}).
Extending these results to a ferromagnet inserted in
a Josephson junction by the correspondence of
Eq.(\ref{heff}), we find that the new
spin wave dynamics leads to 
the supercurrent, 
\begin{eqnarray}
\langle I(t) \rangle = \sin \phi(t)  \int\frac{d^{d}q}{(2 \pi)^{d}} 
\Big[2 \pi^{2} e \rho^{2} \Delta(|T_{0}|^{2} 
- \frac{3}{4} |T_{1}|^{2})  \nonumber
\\ + 4e |T_{1}|^{2} \rho^{2} h_{eff}(\vec{q}) 
\langle S_{z}(\vec{q}) \rangle \Big],
\label{I_quad}
\end{eqnarray} 
with $\rho$ the spin density of states within the leads.
A matching of the Josephson and spin frequencies
(such as present here for variations
in the low temperature magnetic 
moment (see Eqs.(\ref{deltaSz},\ref{deltaSzq}))) 
leads to a DC signal; additional
harmonics further appear.
We emphasize that in the above we compared 
only the Gaussian portion in the Bose fields.
Higher order (non-Gaussian) terms originating from Eq.(\ref{dm})
as well as phasons alter the natural correspondence of Eq.(\ref{heff}). 
A full discussion of these issues will be detailed elsewhere \cite{long}.

\section{Other Geometries}
\label{Geometry.}

If phason
contributions are neglected, then by a trivial change of geometry all
of our results thus far, will apply
for other systems as well. For instance, by replacing one of
the superconducting leads
by a surface, the resulting
system may emulate a superconducting
tip coupled to superconducting surface
through a single spin or a ferromagnet.
Here, all of the results of Sections(\ref{single_spin},\ref{ferroS}) 
for the spin dynamics and tunneling current hold.

Similarly, by replacing both superconducting
leads by surfaces and examining a magnetic
layer inserted in between, the resultant system
looks much alike a layered superconducting/magnetic 
system. In this system, the results of 
Section(\ref{ferroS}) apply.

\section{Large $S$ Adiabatic Approximations}
\label{Large_S.}

Thus far we studied the 
dynamics of single spins and 
of ferromagnets. In \cite{nut03},
the large $S$ limit of the
single spin problem was
studied. In that work, 
several approximations
were made: 

(i) The perturbative approach that
we employed in the current article 
which allows an exact
evaluation of all pertinent integrals
to low orders was replaced
by an ``adiabatic'' approximation
wherein the slow dynamics
of the spin vis a vis electronic
processes was explicitly incorporated,
$\vec{S}(t^{\prime}) \simeq \vec{S}(t) + (t^{\prime} - t) 
(d\vec{S}/dt)$.

(ii) The (``classical'') large $S$ limit allowed us to
omit many instances of $\vec{S}_{qu}$ in the equations
of motion and only $\beta^{R}$ related contributions
in the tunneling action were consequential. Furthermore,
as briefly alluded to earlier, in this limit, 
the average of the vectorial product
$\langle \vec{S}(t^{\prime}) \times \vec{S}(t) \rangle$
is equal to the product of the averages $\langle \vec{S}(t^{\prime}) \rangle
\times \langle \vec{S}(t) \rangle$. Correspondingly, any expectation
value braces may be omitted. Thus,  
we may replace any expectation value $\langle {\cal{A}} \rangle$ 
by ${\cal{A}}$ itself.

The advantage of this method is that
furnishes an elegant non-perturbative closed 
form solution for the spin dynamics.
We will not repeat the results
for the single spin cluster
$(S \gg 1 )$ problem here and rather
refer the reader to \cite{nut03}.
We now briefly comment on applications
of this method to other systems.

To ${\cal{O}}(|T_{1}|^{2})$,
the spin wave dynamics
in ferromagnets
may be attained via the 
substitution of Eq.(\ref{heff}).
Equivalently, the spin wave equations
of motion may also be determined directly
when applying the adiabatic approximation
on Eq.(\ref{class_final1}).
We then find
\begin{eqnarray}
\frac{d\vec{S}_{cl}(\vec{r}_i)}{dt} + 
\vec{h} \times \vec{S}_{cl}(\vec{r}_{i}) 
+ \sum_{j} J_{ij} \vec{S}_{cl}(\vec{r}_{j},t)  \times 
\vec{S}_{cl}(\vec{r}_{i},t) \nonumber
\\ + \kappa \vec{S}_{cl} 
\times \frac{d \vec{S}_{cl}}{dt} 
\sin \omega_{J} t =0
\end{eqnarray}
with  $\kappa \equiv \sum_{k,p} \frac{|\Delta|^{2} |T_{1}|^{2}}{E_{k} E_{p}}
[(E_{k}+E_{p}-eV)^{-2}-(E_{k}+E_{p}+eV)^{-2}]$.
The appropriate spin wave equation is 
\begin{eqnarray}
\frac{d b(\vec{q})}{dt} = i[h+S\{J(\vec{q}) - J(0) \}] 
b(\vec{q}) \nonumber
\\ + \kappa \partial_{t} b(\vec{q},t) \sin \omega_{J} t.
\label{swk}
\end{eqnarray}
The solution to Eq.(\ref{swk}) is 
\begin{eqnarray}
b(\vec{q}, t) = b(\vec{q},0) \exp 
[ - \frac{2i(S\{J(\vec{q}) - J(0) \}
+h)}{\omega_{J} \sqrt{1 - \kappa^{2}}} \nonumber
\\ \times 
\{ \tan^{-1} (\frac{\kappa}{\sqrt{1-\kappa^{2}}}) - \tan^{-1}(\frac{\kappa- 
\tan(\omega_{J} t/2)}{\sqrt{1 - \kappa^{2}}})\}],
\label{nonuniform}
\end{eqnarray}
which is quite different from the standard spin-wave evolution
in a magnet outside a Josephson junction.
The key feature is a nonuniform evolution
of each spin-wave. Similar to the azimuthal precession
of a single spin, the planar components $S_{x,y}$
precess as the real and imaginary parts of $\exp[i \varphi(t)]$
with a nonlinear $\varphi(t)$. Thermodynamic
quantities computed via the corrected bosonic spin-wave dispersion
exhibit corrections.

Similarly, we may examine the adiabatic large $S$ equations of motion 
for an antiferromagnetic spin chain oriented along the z-axis
in a Josephson junction (just as in Fig.(\ref{FIG:MAGNET}) yet
now with a single antiferromagnetic 
spin chain replacing the ferromagnetic slab
in an $h=0$ background). We then find that that
the staggered spin, $\vec{\tilde{S}}_{i} \equiv (-1)^{i} \vec{S}_{i}$
(with the integer $i$ the spin site location along the chain) satisfies
\begin{eqnarray}
0= \Box \vec{\tilde{S}}_{cl}(t) + 3 \partial_{t} \vec{\tilde{S}}_{cl}(t) 
\times \partial_{z} \vec{\tilde{S}}_{cl}(t) \nonumber
\\ +  \kappa 
\partial_{t} \vec{\tilde{S}}_{cl}(t) \sin \omega_{J} t,
\label{theta}
\end{eqnarray}
where
$\Box \equiv 
\frac{v_{s}}{g} [ \partial_{z}^{2} - \frac{1}{v_{s}^{2}} 
\partial_{t}^{2}]$.
Here,  $g=2/S$ and the spin wave velocity $v_{s} = 2aJS$,
with $a$ is the lattice constant.
The role of the supercurrent
as an effective driving term is evident in the 
last line of Eq.(\ref{theta}).

\section{Conclusions}
\label{conc.}

In conclusion, our work addresses new dynamical
effects exhibited by spins in Josephson Junctions.
En route, many features (general and specific) were
found:
\bigskip{ }

{\bf(1)} We derived the 
{\em exact} equation of motion for
spin systems on Keldysh contours.

{\bf(2)} The $S=1/2$ spin dynamics
of a single spin in a Josephson junction
was investigated and a perturbative
solution was given. {\em Spin-1/2 Josephson
nutations are predicted.} 

{\bf(3)} Spin dynamically triggered ac effects
are predicted.

{\bf(4)} The spin wave dynamics {\em of a ferromagnet
in between two superconducting leads} was
investigated. We predict non-trivial
spin wave dynamics as well as
new manifestations of
this dynamics (most
notably in the supercurrent).

{\bf(5)} Large $S$ expressions were discussed
for ferromagnetic slabs and antiferromagnetic
spin chains in a Josephson junction.

\bigskip{ }

{\bf Acknowledgments}

This work was supported by the US DOE under LDRD X1WX (ZN and AVB).
We gratefully acknowledge discussions with Yu Makhlin.

\subsection{Appendix A: Detection}

The non-trivial spin-wave and associated supercurrent 
in Josephson junctions containing ferromagnets
(section \ref{ferroS}) may be seen more readily seen 
than those of single spins. The spin dynamics
may be discerned by measuring the 
magnetization of the ferromagnetic
slab as a function of time (as
suggested by Eq.(\ref{deltaSzq})) as well
as by monitoring the supercurrent
(given by Eq.(\ref{I_quad})).
Other techniques may involve standard
measurements of microwave radiation from the junction
(and backaction effects). The
magnitudes of these effects will be 
studied elsewhere \cite{long}.

We now briefly review a detection scheme discussed in \cite{nut03},\cite{snz} 
for the Josephson nutations for the $S \gg 1$ limit of 
the general spin S results of Section(\ref{retS}).
This corresponds to a single magnetic cluster.

As it moves, the spin cluster magnetic moment generates a time-dependent magnetic field, $\delta{\vec B}({\bf 
r},t) = \frac{\mu_{0}}{4 \pi} [3 {\vec r} ({\vec r} \cdot {\vec m}(t))
-   r^{2} {\vec m}(t)]/r^{5}$. This small field is superimposed against
the constant external field background (${\vec B}$).
In the above, ${\vec r}$ is the position relative to the spin and
${\vec m}(t)$ is the magnetic moment of the spin.
A ferromagnetic cluster of spin $S = 100$ generates a detectable magnetic field 
$\delta B \sim 10^{-10}$~T at a distance of a micron
away from the spin. For a SQUID loop of micron dimensions located 
at that position, the ensuing flux variation 
$\delta \Phi \sim 10^{-7} \Phi_{0}$ (with
$\Phi_{0} = hc/e$ the flux quantum) are within reach of modern SQUIDs. In
such a setup, with $T_{1}/T_{0} \sim 0.1$, the typical critical
Josephson current
is $J_{S}^{(0)} \sim 10 \;\mu\mbox{A}$,
$\vert\Delta\vert=1\;\mbox{meV}$, and $eV \sim 10^{-3}
|\Delta|$, we find that the relative corrections $ \delta S/S \sim 0.1$. The spin
components orthogonal to ${\vec B}$ vary, to ${\cal{O}}(T_{1}^{2})$, 
with Fourier components at frequencies $|\omega_{L} \pm
\omega_{J}|$ ($\omega_L= g \mu_{B} B$), leading to a observable signal in the magnetic
field ${\vec B} + \delta {\vec B}$.  For a field $B \sim 200$~G, $\omega_{L} \sim 
560\; \mbox{MHz}$, and a new side band will
appear at $|\omega_{L} - \omega_{J}|$, whose magnitude may be tuned
to 10--100~MHz. This measurable frequency is
easily distinguished from the Larmor frequency $\omega_{L}$.

The efficiency of this detection scheme may be enhanced by  
embedding the spin in one of the Josephson junction arms of the 
SQUID itself. Such a setup is illustrated in Fig.~\ref{FIG:DETECTION}.
The Josephson junction harboring the spin is employed
in both triggering the nutations and,  along with the second 
junction of the SQUID, in the detection of the resulting nutations. 
\begin{figure}
\centerline{\includegraphics[width=0.8 \columnwidth]{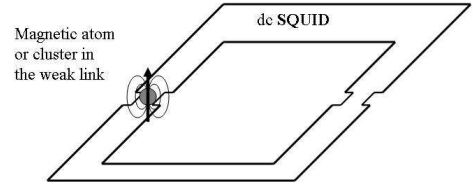}}
\caption{A SQUID-based detection scheme. The SQUID monitors the magnetic field
produced by the magnetic cluster in one of the junctions.}
\label{FIG:DETECTION}
\end{figure}

\subsection{Appendix B: Time Ordering Along the Keldysh 
Contour}

In averaging within the path integral formalism, 
we immediately attain time ordered averages. 
In interchanging the order of the spins (if necessary)
in the vectorial product upon time ordering
within the path integral $CP_{1}$ 
formulation a change of sign is incurred \cite{explain}. 
We now go over, in some detail, time ordering 
within the Keldysh framework. As the time
ordering is performed along the Keldysh 
contour, we will denote it by $T_{K}$
as we have done in deriving the 
effective action of subsection(\ref{eff_action}).

Consider the third term in
Eq.(\ref{class_final}). Upon time ordering,
we find that
\begin{eqnarray}
\langle \vec{S}_{qu}(t^{\prime}) \times \vec{S}_{cl}(t) \rangle_{S}  = 
\langle T_{K} [\vec{S}_{qu}(t^{\prime}) \times \vec{S}_{cl}(t)]
 \rangle \nonumber
\\ = \langle T_{K}[ \frac{1}{2}(\vec{S}_{up}(t^{\prime}) 
- \vec{S}_{down}(t^{\prime})) \nonumber
\\ \times
(\vec{S}_{up}(t) + \vec{S}_{down}(t))] \rangle.
\label{exampleorder}
\end{eqnarray}
Due to the form of the Keldysh contour 
(see Fig.(\ref{FIG:keldysh1})), irrespective
of the values of $t$ and $t^{\prime}$,
$\vec{S}_{up}(t)$ always appears before $\vec{S}_{down}(t^{\prime})$.
Similarly, for $t>t^{\prime}$, $\vec{S}_{up}(t)$ appears after
$\vec{S}_{up}(t^{\prime})$ while $\vec{S}_{down}(t)$ appears before
$\vec{S}_{down}(t^{\prime})$.
With this information at hand, 
Eq.(\ref{exampleorder}) leads
to 
\begin{eqnarray}
\langle T_{K} [ \vec{S}_{qu}(t^{\prime}) \times 
\vec{S}_{cl}(t)] \rangle \nonumber
\\  = - \theta(t- t^{\prime}) [\langle \vec{S}(t^{\prime}) 
\times \vec{S}(t) 
\rangle + \langle \vec{S}(t) \times \vec{S}(t^{\prime}) \rangle].
\end{eqnarray}
The expectation values on the right are the usual operator 
expectation values. Here, we disposed of the up/down indices 
once we took care
of time ordering. The up/down labels merely
serve as mnemonics for this time ordering
along the Keldysh contour. 

Similarly, we find that
\begin{eqnarray}
\langle T_{K} 
[\vec{S}_{cl}(t^{\prime}) \times \vec{S}_{cl}(t)] \rangle = \nonumber
\\ \frac{1}{2} [\theta(t^{\prime}-t) (\langle
\vec{S}(t^\prime) \times \vec{S}(t)
- \vec{S} (t) \times \vec{S} (t^{\prime})) \rangle \nonumber
\\ + \theta(t-t^{\prime}) ( \langle \vec{S} (t)
\times \vec{S}(t^{\prime})
- \vec{S} (t^\prime) \times \vec{S}(t)) \rangle ].
\label{t1}
\end{eqnarray}
By the same token, 
$\langle T_{K} [\vec{S}_{qu}(t) \times \vec{S}_{qu}(t^{\prime})] \rangle = 0.$

As will become clear shortly, in the solution of Eqs.(\ref{class_final}) to
order ${\cal{O}}(T_{1}^{2})$, we will need the spin-spin expectation values
of the usual Larmor problem (i.e. a single spin
in a magnetic field sans any supercurrent). 
Here, 
\begin{eqnarray}
S_{x}(t) = S_{x}(0) \cos \omega_{L} t + S_{y}(0) \sin \omega_{L} t \nonumber
\\ S_{y}(t) = S_{y}(0) \cos \omega_{L}t 
- S_{x}(0) \sin \omega_{L} t, \nonumber
\\ S_{z}(t) =  S_{z}(0),
\label{Larmoreq}
\end{eqnarray}
with the external magnetic field 
oriented along the positive z axis 
and $\omega_{L} = |\vec{h}|$ the Larmor frequency.

Next, we invoke this solution to compute the various 
expectation values within the Larmor problem (i.e.
to order ${\cal{O}}(T_{1}^{0})$). We find that
\begin{eqnarray}
\langle \vec{S}_{qu}(t^{\prime}) \times \vec{S}_{cl}(t) \rangle_{S} \nonumber
\\ =
- \theta(t - t^{\prime}) 
[i ~ Im \{ \langle \vec{S}(t^{\prime}) \times \vec{S}(t) \rangle \} \nonumber
\\ = -i \theta(t - t^{\prime}) \{[\langle S_{x}(t) \rangle
(1+ \cos \omega_{L}(t^{\prime} -t)  ) \nonumber
\\ 
+ \langle S_{y}(t) \rangle \sin \omega_{L}(t^{\prime} -t)] 
\hat{e}_{x} \nonumber
\\ + [ \langle S_{y}(t) \rangle (1+ \cos \omega_{L} (t^{\prime} - t)) 
\nonumber
\\ - 
\langle S_{x}(t) \rangle \sin \omega_{L}(t^{\prime} - t)] 
\hat{e}_{y} \nonumber
\\ + 2 \langle S_{z}(t) \rangle \cos \omega_{L}(t^{\prime} - t) 
\hat{e}_{z}\}.
\label{Ixy}
\end{eqnarray}

Similarly, for the $S=1/2$ problem, 
\begin{eqnarray}
\langle \vec{S}_{cl}(t^{\prime}) \times \vec{S}_{cl}(t) \rangle_{S} \nonumber
\\  =
Re\{ \langle \vec{S}(t^{\prime}) \times \vec{S}(t) \rangle \} \nonumber
\\ = 
\frac{1}{2} \sin \omega_{L}(t^{\prime} - t) \hat{e}_{z}.
\label{Iz}
\end{eqnarray}

In Eqs.(\ref{Ixy},\ref{Iz}), we vividly see that
upon time ordering along the Keldysh contour,
the non-vanishing 
spin cross products become simply related to the imaginary and real
parts of $\langle \vec{S}(t^{\prime}) \times \vec{S}(t) \rangle $.

\subsection{Appendic C: Evaluation of Integrals}
\label{integrals}

We are now ready for the evaluation of the various integrals $I$ that 
appear in Eq.(\ref{class_final}) to order ${\cal{O}}(T_{1}^{2})$.

We start with $\vec{I}_{cl-cl}$. Inserting Eq.(\ref{Iz}) and Eq.(\ref{BetaR})
into Eq.(\ref{class_final}) we find upon invoking 
the relation $\phi(t) = \omega_{J} t$
\begin{eqnarray}
\langle \vec{I}_{cl-cl} \rangle_{S} =  \nonumber
\\ 4 |T_{1}|^{2} \hat{e}_{z}
\int dt^{\prime} j(t,t^{\prime})
\beta^{R}(t,t^{\prime}) \langle \vec{S}_{cl}(t^{\prime})
\times \vec{S}_{cl}(t) \rangle_{S} \nonumber
\\ = - |T_{1}|^{2} \hat{e}_{z} \sum_{k,p} \frac{\Delta^2}{E_k E_p}
\int dt^{\prime} \theta(t-t') \cos \omega_{J} 
\frac{t+t^{\prime}}{2} \nonumber
\\ \times \sin \omega_{L}(t - t^{\prime}) \sin [(E_k+E_p)(t-t')].
\label{intbr}
\end{eqnarray}

Before evaluating Eq.(\ref{intbr}) exactly, we illustrate
what answer is anticipated.
The underlying observation of this {\em adiabatic} approach
is that, as a consequence of $\omega_{L,J} \ll E_{k,p}$
the spin dynamics is far slower than that of electronic
processes. Thus, in integrals involving both 
spin and electronic degrees of freedom, we 
may regard the spin as nearly stationary and 
approximate $\vec{S}(t^{\prime}) \simeq \vec{S}(t) + (t^{\prime} - t) 
(d\vec{S}/dt)$. This physically transparent approximation was 
invoked in \cite{nut03}. Employing this approximation 
here we anticipate that 
$\langle \vec{I}_{cl-cl} \rangle_{S} \simeq C_{1}  \hat{e}_{z} 
\sin \omega_{L} t$
where $C_{1}$ is, up to trivial prefactors, given by 
$\int_{0}^{\infty} d x~  [x^{2} \beta^{R}(x)]$.
Such an anticipation is not far off the mark.

Next, we exactly evaluate Eq.(\ref{intbr}) by rewriting products 
of trigonometric functions as sums and consequently employing
the identities 
\begin{eqnarray}
\int_{0}^{\infty} dx ~\cos a x = \pi \delta(a), \nonumber
\\ \int_{0}^{\infty}  dx ~ \sin ax = \pi \delta(a) + \frac{1}{a}.
\label{identityint}
\end{eqnarray}

In the integrals of interest, $x$ assumes the role of $(t^{\prime}-t)$.
As the applied magnetic field and voltage are far lower
than electronic energy scales, $\omega_{L,J} \ll E_{k,p}$, 
we find that all resonances signaled 
by the delta functions are physically unaccessible and our 
expressions undergo further simplifications. Retaining the leading
order terms in ${\cal{O}}(\omega_{L,J}/(E_{k}+E_{p}))$ we arrive at
\begin{eqnarray}
\langle \vec{I}_{cl-cl} \rangle_{S}  = - |T_{1}|^{2} \hat{e}_{z} \sum_{k,p} 
\frac{\Delta^{2} \omega_{L} \omega_{J}}{E_{k} E_{p} (E_{k} + E_{p})^{3}} 
\sin \omega_{J}t.
\label{Icl-cl*}
\end{eqnarray}
Thus, the form anticipated by the adiabatic approximation
is correct if $C_{1} = - |T_{1}|^{2} \sum_{k,p} 
\frac{\Delta^{2} \omega_{L} \omega_{J}}{E_{k} E_{p} (E_{k} + E_{p})^{3}}$.

Similarly, by inserting Eqs.(\ref{BetaK},\ref{Ixy}) 
into Eq.(\ref{class_final}) and invoking Eqs.(\ref{identityint}), 
we find
\begin{eqnarray}
\langle \vec{I}_{qu-cl} \rangle_{S}  =  - |T_{1}|^{2}
 \sum_{k,p} \frac{|\Delta|^{2}}{E_{k} E_{p} (E_{k}+E_{p})^{2}} \nonumber
\\ \times
[(\langle S_{x}(t) \rangle \omega_{J} \sin \omega_{J} t - 
\langle S_{y}(t) \rangle \omega_{L} \cos \omega_{J} t) \hat{e}_{x} \nonumber
\\ + (\langle S_{y}(t) \rangle \omega_{J} \sin \omega_{J} t
+ \langle S_{x}(t) \rangle \omega_{L} 
\cos \omega_{J} t) \hat{e}_{y} \nonumber
\\ + \langle S_{z}(0) \rangle \omega_{J} \sin \omega_{J} 
t \hat{e}_{z}] \nonumber
\\ = C_{2} [\omega_{J} \langle \vec{S}(t) \rangle 
\sin \phi(t) + \omega_{L} (\hat{e}_{z} \times 
\langle \vec{S}(t) \rangle) \cos \phi(t)],
\label{Iqu-cl*} 
\end{eqnarray}
with the constant $C_{2} \equiv  - |T_{1}|^{2}
\sum_{k,p} \frac{|\Delta|^{2}}{E_{k} E_{p} (E_{k}+E_{p})^{2}}$
and $\phi(t) = \omega_{J} t$ the superconducting phase
difference across the junction. The last line
of Eq.(\ref{Iqu-cl*}) has a very physically suggestive
meaning regarding spin contractions and
an effective longitudinal magnetic 
field- items which we will expand on in  
Section(\ref{consequences}).
Our expressions (Eqs.(\ref{Icl-cl*},\ref{Iqu-cl*})) 
above are exact to lowest order in $T_{1}$ and the ratios
$(\omega_{L,J}/E_{k,p})$.

Finally, the integral $\langle \vec{I}_{qu-qu} \rangle =0$ 
identically by virtue 
of a vanishing
$\langle \vec{S}_{qu}(t^{\prime}) \times \vec{S}_{qu}(t) \rangle_{S} =0$.


\begin{references}
\bibitem{explain_CP1} For a coherent state oriented along 
$\hat{\Omega} = (\sin \theta \cos \phi, \sin \theta \sin \phi, 
\cos \theta)$ on the unit sphere, the corresponding 
spinor is, up to an arbitrary multiplicative global phase 
factor ($\exp[i \chi]$), the spinor
\begin{eqnarray}
z^{*}= (\cos \frac{\theta}{2} e^{-i \phi/2}, 
\sin \frac{\theta}{2} e^{i \phi/2}).
\label{ZCP1.}
\end{eqnarray} 
While inserting complete coherent spin states at all times
steps we invoke 
\begin{eqnarray}
\langle \Omega | \Omega^{\prime} \rangle = (\frac{1 + \hat{\Omega} \cdot 
\hat{\Omega}^{\prime}}{2})^{1/2} e^{-i S \psi}
\nonumber
\\ \psi = 2 \tan^{-1}\Big( 
 \tan[\frac{1}{2}(\phi - \phi^{\prime})]  \frac{\cos[\frac{1}{2}
(\theta + \theta^{\prime})]}{\cos[\frac{1}{2}(\theta - \theta^{\prime})]} 
\Big) 
+ \chi - \chi^{\prime}
\end{eqnarray}
where $\chi$ and $\chi^{\prime}$ depend on gauge convention,
with $| \Omega \rangle$ and $| \Omega' \rangle$
spin coherent states at two different times.
In the limit of small deviations $(\phi^{\prime} - \phi)$
and $(\theta^{\prime} - \theta)$, the above argument
reduces to  $d \psi = [(d \phi/dt) \cos \theta] dt$
which is simply the area elements swept by $\hat{\Omega}(t)$
in an infinitesimal time increment $dt$. Within the path integral
formulation we insert complete sets of coherent states
at different times. We see from the internal product
above that within the spin coherent basis the path integral attains a
Berry phase which is simply the area swept by the spin
its cyclic evolution. (It is cyclic due to the requirement of taking a 
trace within the partition function). 

An explicit evaluation with the spinor
of Eq.(\ref{ZCP1.}) reveals that 
\begin{eqnarray}
{\cal{L}}_{Berry} = z_{a}^{*} (i \partial_{t}) z_{a}
\end{eqnarray}
is none other than $[(i/2) (d \phi/dt) \cos \theta]$. Upon integration
over time this leads, up to trivial additive constants, 
to $(i/2)$ times the area spanned by the spin.
\bibitem{explain}  The illustration of this 
change of sign upon exchanging the order 
of times within the
path integral average is straightforward. 
\begin{eqnarray}
\int Dz Dz^{*} \delta(|z|^{2}-1) \epsilon_{ijk} 
z^{*}_{a}(t_{2}) \sigma^{i}_{ab} z_{b}(t_{2}) 
z^{*}_{c}(t_{1}) \sigma^{j}_{cd} z_{d}(t_{1}) e^{iS} = \nonumber
\\ - \int Dz Dz^{*} \delta(|z|^{2}-1) \epsilon_{ijk} 
z^{*}_{a^{'}}(t_{1}) \sigma^{i}_{a^{'}b^{'}} z_{b^{'}}(t_{1}) 
z^{*}_{c^{'}}(t_{2}) \sigma^{j}_{c^{'}d^{'}} z_{d^{'}}(t_{2}) e^{iS}. \nonumber
\end{eqnarray} 
This can be seen by making an identification
of summation indices, $a^{'} =c,~ b^{'} = d, ~c^{'} = a,~ d^{'} =b$,
and invoking $\epsilon_{ijk} = - \epsilon_{jik}$.
\end{references}
\end{document}